\documentclass[12pt]{article}
\pdfoutput=1
\usepackage{epsf,amsfonts,amssymb,epsfig,amsmath,mathtools,graphics,slashed}
\usepackage{hyperref,hep,graphicx,subfig}
\usepackage{rotating}

\newcommand{\nn}{\notag \\}

\makeatletter

\@addtoreset{equation}{section}
\makeatother

\begin{document}

\begin{titlepage}

\vfill

\begin{flushright}
Imperial/TP/2017/JG/05\\
DCPT-17/29
\end{flushright}

\vfill

\begin{center}
   \baselineskip=16pt
   {\Large\bf Diffusion for Holographic Lattices}
  \vskip 1.5cm
  \vskip 1.5cm
Aristomenis Donos$^1$, Jerome P. Gauntlett$^2$ and Vaios Ziogas$^1$\\
     \vskip .6cm
     \begin{small}
      \textit{$^1$Centre for Particle Theory and Department of Mathematical Sciences\\Durham University, Durham, DH1 3LE, U.K.}
        \end{small}\\    
         \begin{small}\vskip .6cm
      \textit{$^2$Blackett Laboratory, 
  Imperial College\\ London, SW7 2AZ, U.K.}
        \end{small}\\
        \end{center}
     \vskip .6cm
\vfill

\begin{center}
\textbf{Abstract}
\end{center}
\begin{quote}
We consider black hole spacetimes that are holographically dual to strongly coupled field theories
in which spatial translations are broken explicitly.
We discuss how the quasinormal modes associated with diffusion of heat and charge can be systematically constructed in a long wavelength perturbative expansion. We show that the dispersion relation for these modes is given in terms of the 
thermoelectric DC conductivity and static susceptibilities of the dual field theory and thus we derive a generalised Einstein relation from Einstein's equations.
A corollary of our results is that thermodynamic instabilities imply specific types of dynamical instabilities of the associated black hole solutions.
\end{quote}

\vfill

\end{titlepage}

\setcounter{equation}{0}
\section{Introduction}

Holography provides a powerful theoretical framework for studying the properties of strongly coupled quantum critical systems.
A basic feature is that a given quantum system in thermal equilibrium is described by a stationary black hole spacetime with a Killing horizon and,
furthermore, the entropy and conserved charges can be universally determined by data on the black hole horizon (e.g. see \cite{Papadimitriou:2005ii} and references therein). 
Going beyond thermal equilibrium and moving to the realm of linear response, it has been shown that the thermoelectric DC conductivity, when finite, is also universally determined by data on the horizon, by solving a specific Stokes flow for an auxiliary fluid on the horizon \cite{Donos:2015gia}
(further extensions are discussed in \cite{Banks:2015wha,Donos:2015bxe,Donos:2017oym,Donos:2017mhp}). 

It is natural to enquire if other properties of the dual field theory can also be obtained from horizon data.
In this paper we discuss the construction of quasi-normal modes 
that are dual to the long wave-length hydrodynamic modes associated with diffusion of heat and electric charge. In particular, we will show how the dispersion relation for these modes can also be obtained in terms of the properties of the black hole solutions at the horizon.

Recall that in the specific context of translationally invariant and charge neutral systems the diffusion of electric charge was first discussed some time ago
in \cite{Kovtun:2003wp}. Furthermore, again for this specific setup, an Einstein relation, relating the associated electric diffusion
constant to the finite DC conductivity and static charge susceptibilities, was 
derived in \cite{Iqbal:2008by}, where it was also shown how the DC conductivity can be obtained explicitly from the horizon\footnote{From 
the universal perspective of \cite{Donos:2015gia}, this is a special set up where the Stokes flow equations are solved trivially.}. It should
be noted that in this set up the thermal DC conductivity is infinite and, correspondingly, there is
no heat diffusion mode. In this paper we will discuss the diffusion of both electric charge and heat within the general context of 
charged and spatially inhomogeneous media. The spatial inhomogeneities that we consider
arise from breaking of spatial translations explicitly, and the black holes are known as `holographic lattices' \cite{Horowitz:2012ky}. 

In a recent paper \cite{Donos:2017gej} we carried out an analysis of the diffusion of conserved charges in the context of spatially inhomogeneous media for arbitrary quantum field theories (not necessarily holographic). Subject to the retarded current-current correlators satisfying some general analyticity conditions,
as well as assuming that the thermoelectric DC conductivity is finite, the long wavelength hydrodynamical modes associated with diffusion of charge and heat were identified and a generalised Einstein relation was derived. In addition, the general formalism was illustrated for thermoelectric diffusion within the context of relativistic hydrodynamics where momentum dissipation was achieved not by modifying the conservation equations, as is usually done, but by explicitly breaking translations by considering the system with spatially modulated sources for the stress tensor and electric current as in \cite{Banks:2016wdh,Scopelliti:2017sga}.

Given the general results presented in \cite{Donos:2017gej}, one anticipates that it should be possible to derive the generalised Einstein relations 
within the context of general classes of holographic lattices. Specifically, we will consider cases in which 
the DC conductivity is finite\footnote{In particular, we will not be considering superfluids.} and equal to a horizon DC conductivity that is obtained by solving a Stokes flow on the horizon.
One can then ask about the relevant charge susceptibilities.
Since the conserved charges can be evaluated at the horizon 
provided one knows how this data depends on changing the temperature and the chemical potential of the black holes in thermal equilibrium, one can also obtain horizon expressions for the susceptibilities. As we will see, this simple observation about the susceptibilities will be sufficient to
extract the dispersion relations for the diffusive modes and hence the Einstein relation. In slightly more detail, using a radial decomposition of the
equations of motion, we will explain how the quasi-normal diffusion modes can be systematically constructed in a long wavelength, perturbative expansion. In general, while both the radial equations and the constraint equations are required to carry out this construction, we will see that an analysis of just the constraint equations on the horizon are sufficient to extract the Einstein relation,
which is the universal part of the dispersion relation in the long wavelength expansion for the diffusive modes.

Recently there has been a particular focus on studying diffusion of heat and charge in the context of holography. 
This stems, in part, from the suggestion that diffusive processes may be a key to understanding universal aspects of
transport in incoherent metals \cite{Hartnoll:2014lpa}. Furthermore, it was also suggested in \cite{Hartnoll:2014lpa} that
there might be lower bounds on diffusion constants by analogy with bounds on shear viscosity associated with diffusion of momentum \cite{Kovtun:2004de}. A key idea is to write $D\sim v^2\tau$, where $D$ is suitable diffusion constant
and $v$, $\tau$ are characteristic velocities and time scales of the system, and it was suggested in 
\cite{Hartnoll:2014lpa} that $\tau$ should be the `Planckian time scale' $\tau= \hbar/(k_BT)$ \cite{Damle:1997rxu,zaanenhigh}. 
An interesting subsequent development was the suggestion that $v$ should be identified with the butterfly velocity, $v_B$, extracted from
out of time order correlators \cite{Blake:2016wvh,Blake:2016sud} and used as a measure of the onset of quantum chaos.

While there has been a range of interesting holographic results in this direction, including 
\cite{Lucas:2016yfl,Ling:2016ibq,Blake:2016jnn,Davison:2016ngz,Baggioli:2016pia,Wu:2017mdl,Kim:2017dgz,Baggioli:2017ojd,Blake:2017qgd,Wu:2017exh,Giataganas:2017koz, Grozdanov:2017ajz, Lucas:2017ibu,Donos:2017sba}, with an appreciation that it is the 
thermal diffusion should be related to $v_B$, it is fair to say that within holography a sharp global picture has yet to emerge\footnote{It is striking
that a relation of the form 
$D\sim v^2_B\tau$ has also appeared in a variety of other non-holographic contexts, including \cite{Aleiner:2016eni,Gu:2016oyy,Patel:2016wdy,Bohrdt:2016vhv,Davison:2016ngz,Gu:2017ohj,Chowdhury:2017jzb,Patel:2017vfp,Werman:2017abn}, with
$\tau\sim\lambda_L^{-1}$ where $\lambda_L$ is the Lyapunov exponent \cite{Patel:2016wdy}.}. 
Almost all of the holographic study in this area has been
in the setting of specific types of `homogeneous' holographic lattices \cite{Donos:2013eha,Andrade:2013gsa},
which maintain a translationally invariant metric. In these cases it is straightforward to extract 
$v_B$ by studying a shock wave entering the black hole horizon \cite{Shenker:2013pqa,Maldacena:2015waa} (see also \cite{Sfetsos:1994xa}). A notable 
exception is \cite{Lucas:2016yfl} who studied holographic lattices in one spatial dimension, but working in a 
hydrodynamic, high temperature limit of the background holographic lattice. 
We hope that the present work, which illuminates universal aspects of diffusion for arbitrary spatial modulation
in holography, will be useful in further developments.

In a different direction, our derivation of the dispersion relation leads to a general connection between thermodynamic instabilities and 
dynamic instabilities. 
Some time ago, building on \cite{Gubser:2000ec,Gubser:2000mm}, it was shown in a specific holographic context with a translationally invariant horizon, that thermodynamic instability implies
an imaginary speed of sound, leading to unstable quasi-normal modes and dynamical 
instability\footnote{Some recent discussion of both hydrodynamic and non-hydrodynamic modes and the connection with instabilities in a translationally invariant setting, appeared in \cite{Janik:2015iry,Janik:2016btb,Faedo:2017aoe,Janik:2017ykj}.} \cite{Buchel:2005nt}.
For general spatial modulation within holography, any sound modes will only appear on scales much smaller than the scale of the modulation and hence 
this will not be a universal channel to deduce dynamical instability from thermodynamic instability. Instead, the diffusion modes do provide such a channel.
Specifically, in the presence of spatial modulation, we can deduce the following result.
If the heat and charge susceptibility matrix has a negative eigenvalue, then the system is thermodynamically unstable and then the dispersion relation  implies that there is at least one mixed diffusion mode, involving heat and charge, living in the upper half plane which will necessarily lead to a dynamical instability.

\section{Background black hole solutions}

We will consider a general class of bulk theories which couple the metric to a gauge field $A_\mu$, with field strength $F_{\mu\nu}$, and
a scalar field $\phi$ in $D$ spacetime dimensions, governed by an action of the form
\begin{align}\label{eq:bulk_action}
S=\int d^D x \sqrt{-g}\,\left(R-V(\phi)-\frac{Z(\phi)}{4}\,F^{2}-\frac{1}{2}\left(\partial\phi \right)^2\right)\,.
\end{align}
The only constraints that we impose on the functions $V(\phi),Z(\phi)$ is that the equations of motion admit an $AdS_D$ vacuum solution
with $\phi=A_\mu=0$. We assume that in this vacuum the scalar field $\phi$ is dual to an operator with conformal dimension $\Delta$.
We have also set $16\pi G=1$ for convenience.

We are interested in studying the family of static, background black hole solutions that lie within the ansatz
\begin{align} \label{eq:DC_ansatz}
ds^{2}&=-UG\,dt^{2}+\frac{F}{U}\,dr^{2}+ds^2(\Sigma_d)\,,\nn
A&=a_{t} dt\,,
\end{align}
with $ds^2(\Sigma_d)=g_{ij}dx^i dx^j$ and $d=D-2$. The
functions $G,F,a_t, \phi$ and the metric components $g_{ij}$ are all independent of the time coordinate $t$ and depend on $(r,x^i$).
Note that the function $U=U(r)$, which is redundant, is included to conveniently deal with some aspects of the asymptotic behaviour of the solution. 

Although it is possible to be more general, to simplify the presentation we will assume that we have single black hole Killing horizon, located at $r=0$, and that the coordinates $(t,r,x^i)$ are globally defined outside the black hole
all the way out to the $AdS_D$ boundary which will be located
at $r\to\infty$. In particular, this means that the radial foliation is globally defined up to a
`stretched horizon' located at some small radial distance outside the black hole and that the topology of the black hole horizon is $\Sigma_d$. Similarly, we will also assume $\Sigma_d$ has planar topology
and all functions appearing in \eqref{eq:DC_ansatz} are assumed to be periodic in the spatial directions $x^i$ with period $L_i$, corresponding to static, periodic deformations of the dual CFT. It will be useful to define $\oint =(\prod L_i)^{-1}\int dx^1\dots dx^d$ which allows us to extract the zero mode
of periodic functions.

Asymptotically, as $r\to\infty$, the solutions are taken to approach $AdS_D$ with 
boundary conditions that explicitly break translation invariance:
\begin{align}\label{asymptsol}
&U\to r^2,\qquad F\to 1,\qquad 
G\to G^{(\infty)}(x),\qquad g_{ij}(r,x)\to r^2 g^{(\infty)}_{ij}(x),\nn
&a_{t}(r,x)\to\mu(x),\qquad \phi(r,x)\to  r^{\Delta-d-1} \phi^{(\infty)}(x)\,.
\end{align}
This corresponds to placing the dual CFT on a curved spacetime manifold with metric given by $ds^2=-G^{(\infty)}(x)dt^2+g^{(\infty)}_{ij}(x)dx^idx^j$,
having a spatially dependent chemical potential $\mu(x)$ and deforming by a spatially dependent source $\phi^{(\infty)}(x)$ for the operator dual to $\phi$. It will be convenient to separate out the zero mode of $\mu(x)$ by defining
\begin{align}\label{zmmu}
\mu(x)\equiv \bar\mu+\tilde\mu(x)
\end{align}
with constant $\bar \mu$ and $\oint\tilde \mu(x)=0$. 

The Killing horizon will, in general, be spatially modulated and we have
the following expansions\footnote{We have chosen our coordinates so that possible functions $G^{(0)}(x)=F^{(0)}(x)$ are set to unity.} near $r=0$ 
\begin{align}\label{nhexpbh}
U\left(r\right)&=r\left(4\pi\,T+U^{(1)}\,r+\dots\right)\,,\qquad
G(r,x)=1+G^{(1)}\left(x\right)r+\dots\,,\nn
F(r,x)&=1+F^{(1)}\left(x\right)r+\dots\,,\qquad\qquad\quad
g_{ij}=g_{ij}^{(0)}+g_{ij}^{(1)}r+\dots\,,\nn
a_{t}(r,x)&=r\left(a^{(0)}_{t}+a^{(1)}_{t}\left(x\right)r+\dots\right)\,,\qquad
\phi=\phi^{(0)}(x)+r\phi^{(1)}(x)+\dots\,.
\end{align}
After a Euclidean continuation, we find that the constant $T$ is the Hawking temperature of the black hole
and can be identified with the constant temperature of the dual field theory\footnote{Since the CFT is defined on the boundary metric \eqref{asymptsol} there is also a natural notion of a local temperature of the dual field theory given by $T(x)=[G^{(\infty)}(x)]^{-1/2}T$. Also note that for the case of CFTs we can carry out a Weyl transformation to set $G^{(\infty)}(x)=1$, suitably taking into account the possibility of a conformal anomaly.}.
We also note that we can introduce an ingoing Eddington-Finkelstein-like coordinate 
\begin{align}\label{efdef}
v_{EF}\equiv t+\frac{\ln r}{4\pi T}\,,
\end{align}
and that the metric is regular in the $(v_{EF},r,x^i)$ coordinates as $r\to 0$.

\subsection{Susceptibilities from the horizon}\label{ssusc}
It will be important in the sequel to be able to express certain thermodynamic susceptibilities in terms of data at the horizon. 
More precisely
we can obtain the susceptibilities provided that we know the horizon data as a function of the temperature $T$ and the zero mode
of the chemical potential $\bar\mu$.
We first recall that the total entropy density of the system, $s$, can be expressed as 
\begin{align}
s=4\pi \oint_H \sqrt{g_{(0)}}\,,
\end{align}
where the subscript $H$ emphasises that this is an integral evaluated at the black hole horizon.
Similarly the total charge density, $\rho\equiv J^t$, can be expressed either as a boundary quantity or a horizon quantity via
\begin{align}
\rho\equiv \oint_\infty \sqrt{-g}Z(\phi)F^{tr}= \oint_H \sqrt{g_{(0)}}Z^{(0)}a_{t}^{(0)}\,,
\end{align}
where the equality can be deduced from the gauge equation of motion.

Hence under a constant variation of the temperature, $T\to T+\delta T$, and zero mode 
of the chemical potential, $\bar \mu\to \bar \mu+\delta\bar \mu$ (see \eqref{zmmu}),  
we have 
\begin{align}\label{derivsd}
\delta s\equiv\,& T^{-1}c_{\mu}\,\delta T+\xi\,\delta\bar\mu\,,\nn
\delta \rho \equiv\, &\xi\,\delta T+\chi\,\delta\bar \mu\,,
\end{align}
where the thermodynamic susceptibilities are given by\footnote{In section 3 of \cite{Donos:2017gej}
these quantities were denoted by capital letters:  $C_\mu$, $\Xi$ and $X$.}
\begin{align}\label{susdef}
T^{-1}c_{\mu}&=4\pi \oint_{H} d^{d}x \frac{1}{2}\sqrt{g_{(0)}}\,(g^{(0)})^{ij}\,\frac{\partial g^{(0)}_{ij}}{\partial T}\,,\nn
\xi&=4\pi\,\oint_{H} d^{d}x \frac{1}{2}\sqrt{g_{(0)}}\,(g^{(0)})^{ij}\,\frac{\partial g^{(0)}_{ij}}{\partial \bar \mu}\,,\nn
&=\oint_H d^{d}x\,\sqrt{g_{(0)}}\left(Z^{(0)}\frac{\partial a_{t}^{(0)}}{\partial T}
+\partial_{\phi}Z^{(0)}\,a_{t}^{(0)}\frac{\partial \phi^{(0)}}{\partial T}
+\frac{1}{2}Z^{(0)}\,a_{t}^{(0)}(g^{(0)})^{ij}\,\frac{\partial g^{(0)}_{ij}}{\partial T}\right)\,,\nn
\chi&=\oint_H d^{d}x\,\sqrt{g_{(0)}}\left(Z^{(0)}\frac{\partial a_{t}^{(0)}}{\partial \bar \mu}
+\partial_{\phi}Z^{(0)}\,a_{t}^{(0)}\frac{\partial \phi^{(0)}}{\partial \bar \mu}
+\frac{1}{2}Z^{(0)}\,a_{t}^{(0)}(g^{(0)})^{ij}\,\frac{\partial g^{(0)}_{ij}}{\partial \bar \mu}\right)\,.
\end{align}
The equality of the two expressions for $\xi$ at the horizon is not obvious. However, from a boundary perspective
it is just a Maxwell relation that arises from the first law. To see this we recall that we can calculate the renormalised, total free energy density,
$w_{FE}$, from the total on-shell action after adding suitable boundary terms. For the ensemble of interest we have
$s=-\delta w_{FE}/\delta T$ and $\rho=-\delta w_{FE}/\delta\bar \mu$ and the result at the horizon follows. 
Note that $c_{\mu}\equiv T(\partial s/\partial T)_{\bar\mu}$. Later
we will also need  $c_\rho\equiv T(\partial s/\partial T)_\rho$ which can be written as
\begin{align}\label{ceerho}
c_{\rho}\equiv c_{\mu}-\frac{T\xi^{2}}{\chi}\,.
\end{align}
To see this we use $\xi/\chi=-(\partial\bar \mu/\partial T)_\rho=(\partial s/\partial\rho)_T$, where the second equality is a Maxwell relation.

\section{Time dependent perturbation and the constraints}\label{timedeppert}
Consider a general perturbation of the background black hole solution \eqref{eq:DC_ansatz} given by $\delta P\equiv\{\delta g_{\mu\nu}$, $\delta a_\mu$, $\delta\phi\}$, with all quantities functions of all of the bulk coordinates $(t,r,x^i)$. We want to consider time-dependence 
of the form $e^{-i\omega t}$. It is convenient to write 
\begin{align}\label{peeeq}
\delta P(t,r,x^i)=e^{-i\omega [t+S(r)]}\delta \hat P(r,x^i)\,,
\end{align}
with $S(r)\to 0$ as $r\to\infty $ and, in order to ensure that the perturbation satisfies ingoing boundary conditions at the black hole horizon,
$S(r)\to \frac{\ln r}{4\pi T}+S^{(1)}\,r+\cdots$ as $r\to 0$.

With this in hand, and recalling the definition of $v_{EF}$, the ingoing Eddington-Finkelstein-like coordinate in \eqref{efdef},
we demand that near $r=0$ the perturbation behaves as 
\begin{align}\label{eq:nh_exp1}
\delta g_{tt}&=e^{-i\omega v_{EF}}(4\pi Tr)\left(\delta g^{(0)}_{tt}\left(x\right)+{\cal O}(r) \right),\quad
\delta g_{rr}=e^{-i\omega v_{EF}}\,\frac{1}{4\pi Tr}\,\left( \delta g_{rr}^{(0)}\left(x\right)+{\cal O}(r)\right),\nn
\delta g_{ij}&=e^{-i\omega v_{EF}}\,\delta g_{ij}^{(0)}\left(x\right)+{\cal O}(r),
\qquad\qquad\quad\,
\delta g_{tr}=e^{-i\omega v_{EF}}\,\delta g_{tr}^{(0)}\left(x\right)+{\cal O}(r)\,,\nn
\delta g_{ti}&=e^{-i\omega v_{EF}}\,(\delta g_{ti}^{(0)}\left(x\right)+r\,\delta g_{ti}^{(1)}\left(x\right)+{\cal O}(r^{2})),\nn
\delta g_{ri}&=e^{-i\omega v_{EF}}\,\frac{1}{4\pi Tr}\,\left( \delta g_{ri}^{(0)}\left(x\right)+r\,\delta g_{ri}^{(1)}\left(x\right)+{\cal O}(r^{2}) \right)\,,
\end{align}
as well as 
\begin{align}\label{eq:nh_exp2}
\delta a_{t}&=e^{-i\omega v_{EF}}\,(\delta a_{t}^{(0)}\left(x\right)+r\,\delta a_{t}^{(1)}\left(x\right)+{\cal O}(r^{2})),\nn
\delta a_{r}&=e^{-i\omega v_{EF}}\,\frac{1}{4\pi Tr}\,\left( \delta a_{r}^{(0)}\left(x\right)+r\,\delta a_{r}^{(1)}\left(x\right)+{\cal O}(r^{2}) \right)\,,\nn
\delta a_{i}&=e^{-i\omega v_{EF}}\,(\delta a_{i}^{(0)}\left(x\right)+{\cal O}(r)),\nn
\delta \phi&=e^{-i\omega v_{EF}}\,(\delta \phi^{(0)}\left(x\right)+{\cal O}(r)),
\end{align}
with 
\begin{align}\label{eq:nh_exp3}
-2\pi T(\delta g_{tt}^{(0)}+\delta g_{rr}^{(0)})=-4\pi T\,\delta g_{rt}^{(0)}&\equiv p\,,\notag\\
\delta g_{ti}^{(0)}=\delta g_{ri}^{(0)}&\equiv-v_{i},\notag\\
\delta a_{r}^{(0)}=\delta a_{t}^{(0)}&\equiv w\,.
\end{align}
There is some residual gauge invariance for the perturbation at the horizon,
maintaining the ingoing boundary conditions,
which we discuss in appendix \ref{resgi}.

\subsection{Constraints}\label{coneqsb}
Using a radial decomposition of the equations of motion one obtains a set of constraints that must
be satisfied on a surface of constant $r$. We want to evaluate these constraints for the perturbed solution
at the black hole horizon. More precisely we evaluate the constraints on a stretched horizon located at a small
radial distance $r$ away from the horizon and then take the limit as $r\to 0$. 
The calculations are a generalisation of the calculations that were carried out in \cite{Donos:2015gia,Banks:2015wha}. 
Here we will just state the final result but we have presented some details in appendix \ref{evalcon}.

The combined set of constraints include two scalar equations and a vector equation. If we define
\begin{align}\label{eq:J_hor2}
Q^i_{(0)} &=4\pi T\sqrt{g_{(0)}}v^i\,,\nn
J^i_{(0)}&=\sqrt{g_{(0)}}g^{ij}_{(0)}Z^{(0)}\left(\partial_j w+{a^{(0)}_t}v_j+i\omega \delta a^{(0)}_{j} \right)\,.
\end{align}
then the two scalar equations are
\begin{align}\label{hconstrainta}
\partial_{i} Q^{i}_{(0)}&=i\omega T\Big(2\pi \sqrt{g_{(0)}}g^{ij}_{(0)}\delta g^{(0)}_{ij}\Big)\,,
\end{align}
and
\begin{align}\label{hconstrainttwo}
\partial_{i} J^{i}_{(0)}&=i\omega\sqrt{g_{(0)}}\Big[
Z^{(0)}\left(a^{(0)}_{t}\,\left(\delta g^{(0)}_{tt}+\frac{p}{4\pi T}\right) +\delta a^{(1)}_{t}-\frac{i \omega}{4\pi T}\left( \delta a^{(1)}_{t}-\delta a^{(1)}_{r} \right)\right)
\nn
&\qquad+Z^{(0)} \left(\frac{1}{2}a^{(0)}_{t}g_{(0)}^{ij} \delta g^{(0)}_{ij}  + \frac{1}{4\pi T}v^i \partial_{i}a_{t}^{(0)}\right)
+\partial_{\phi}Z^{(0)}a^{(0)}_{t}\,\delta\phi^{(0)}\Big]\,.
\end{align}
Finally the vector equation can be written as
\begin{align}\label{hiconstraint2b}
&-2\,\nabla^{j}\nabla_{(j}v_{i)}+(1+\frac{i\omega}{4\pi T})\nabla_{i}p -Z^{(0)}a^{(0)}_{t}\,\left( \nabla_{i}w+i\omega\delta a_{i}^{(0)} \right) +\nabla_{i}\phi^{(0)} 
\left( v^{j} \nabla_{j}\phi^{(0)}-i\omega \delta\phi^{(0)}\right)\nn
&\qquad\qquad=i\omega\left(\delta g_{ti}^{(1)}
-\frac{i\,\omega}{4\pi T}(\delta g_{ti}^{(1)}-\delta g_{ri}^{(1)})
+g^{(1)}_{il}v^{l}
-\partial_{i}\delta g_{tt}^{(0)} 
-\nabla^{k}\delta g^{(0)}_{ki} \right)\,.
\end{align}
where the covariant derivative is with respect to the horizon metric $g^{(0)}_{ij}$, which is also used to raise and lower indices.
One can check that these equations are consistent with the residual gauge transformations mentioned above and are given explicitly in 
appendix \ref{resgi}.

Notice that if we set the frequency $\omega=0$ in \eqref{hconstrainta}-\eqref{hiconstraint2b}  
then we precisely recover the Stokes equations derived in \cite{Donos:2015gia,Banks:2015wha}, 
which can be used to obtain the DC conductivity, when finite, after setting the sources in the Stokes equations to zero. 
In fact since these DC Stokes equations with sources will be used later, we record them in appendix \ref{dcapp} for reference.

Also notice that the system of equations does not form a closed set of equations for the perturbation when $\omega \ne 0$. In order to obtain a full solution, we also need to use the radial equations of motion. Interestingly, however, we will show in the next section 
that the constraint equations are sufficient to extract the dispersion relation for the quasinormal diffusion modes. In appendix \ref{contex}
we discuss how the data provided at the horizon and at the $AdS$ boundary allows one, in principle, to solve the full set of Einstein equations.

\section{Constructing the bulk diffusion perturbations}
In this section we explain how one can systematically construct quasi-normal modes that are associated with diffusion 
of heat and charge.  We construct these source-free modes in a long wavelength `hydrodynamic expansion' that is valid for an arbitrary 
background black hole solution \eqref{eq:DC_ansatz}. While the explicit construction of these modes require that one solves 
both the constraint equations at the horizon as well radial equations in the bulk, it is possible to
show that the leading order dispersion relation for the diffusion modes can be expressed 
in terms of the static susceptibilities as well as the `horizon DC conductivity' obtained from a Stokes flow on the horizon given in appendix 
\ref{dcapp}. Now for holographic lattices, when translation invariance is broken explicitly, the
horizon DC conductivity is the same as the DC conductivity of the dual field theory. Thus, our result corresponds to a derivation of an Einstein relation.

We begin by describing a zero mode perturbation that is constructed from thermodynamic considerations. We start with the background ansatz \eqref{eq:DC_ansatz} and then vary the temperature $T$ and the zero mode of the chemical potential $\bar \mu$ via $T+\delta T$ and $\bar \mu+\delta\bar \mu$, where $\delta T$,$\delta\bar\mu$ are real constants. 
This gives rise to a `thermodynamic perturbation' of the metric, gauge field and scalar field of the form 
$\delta g^{TH}_{\mu\nu}=\frac{\partial g_{\mu\nu}}{\partial T}\,\delta T+\frac{\partial g_{\mu\nu}}{\partial \bar \mu}\,\delta \bar\mu$, $\delta A^{TH}=(\frac{\partial a_{t}}{\partial T}\,\delta T+\frac{\partial a_{t}}{\partial  \bar\mu}\,\delta\bar\mu)\,dt$ and $\delta\phi^{TH}=\frac{\partial \phi}{\partial T}\,\delta T+\frac{\partial \phi}{\partial \bar \mu}\,\delta \bar \mu$, respectively. 
By considering the asymptotic behaviour of the background black holes, given in \eqref{asymptsol}, we see that this perturbation has no source
terms for the metric and the scalar, but there is a new source term for the gauge-field of the form $\delta\bar \mu$. As we are interested in source free solutions, we will deal with this in a moment.

We next observe that close to the horizon the above perturbation approaches
\begin{align}\label{nhpeth}
\delta ds^{2}=&-\frac{\delta T}{T}\,\left(4\pi T r\,dt^{2}+\frac{dr^{2}}{4\pi T r} \right)
+\left(\frac{\partial g^{(0)}_{ij}}{\partial T}\,\delta T+\frac{\partial g^{(0)}_{ij}}{\partial \bar \mu}\,\delta \bar \mu\right)\,dx^{i} dx^{j}+\cdots\,,\nn
\delta a_{t}^{TH}&=r\,\left(\frac{\partial a_{t}^{(0)}}{\partial T}\,\delta T+\frac{\partial a_{t}^{(0)}}{\partial \bar \mu}\,\delta \bar\mu\right)+\cdots\,,\nn
\delta\phi^{TH}&=\frac{\partial \phi^{(0)}}{\partial T}\,\delta T+\frac{\partial \phi^{(0)}}{\partial \bar\mu}\,\delta \bar\mu\,.
\end{align}
Notice that this does not satisfy the regularity conditions \eqref{eq:nh_exp1}-\eqref{eq:nh_exp3} required of a real time perturbation. 
To remedy this, and also to remove the extra source term in the gauge field, we perform a time coordinate transformation $t\to t + \frac{\delta T}{T}\,g(r)$ with $g(r)$ vanishing sufficiently fast as $r\to \infty$ and $g(r)=\ln r/(4\pi T)+g^{(1)}r+\dots$ as $r\to 0$. We also perform the gauge transformation $\delta A^{RT}=\delta A^{TH}+d\Lambda$ with $\Lambda=-( t+g(r))\delta \bar \mu$. After performing these transformations we will denote the perturbation with a superscript $RT$ for `real-time'. 

At the horizon this $RT$ perturbation approaches
\begin{align}\label{eq:nh_static_metric_pert}
\delta ds^{2}&=\delta g^{RT}_{\mu\nu}\,dx^{\mu} dx^{\nu}=-\frac{\delta T}{T}\,\left(4\pi T r\,dt^{2}+\frac{dr^{2}}{4\pi T r} \right)-2\,\frac{\delta T}{T}\,dt\,dr\notag\\
&\qquad\qquad\qquad\qquad+\left(\frac{\partial g^{(0)}_{ij}}{\partial T}\,\delta T+\frac{\partial g^{(0)}_{ij}}{\partial \bar \mu}\,\delta \bar\mu\right)\,dx^{i} dx^{j}+\cdots\,,
\nn
\delta a_{t}^{RT}&=-\delta\bar \mu+r\,\left(\frac{\partial a_{t}^{(0)}}{\partial T}\,\delta T+\frac{\partial a_{t}^{(0)}}{\partial \bar\mu}\,\delta \bar\mu\right)+\cdots\,,\nn
\delta a_{r}^{RT}&=-\delta \bar \mu\,(4\pi T\,r)^{-1}+\frac{\delta T}{T}(4\pi T)^{-1}a_{t}^{(0)}-g^{(1)}\delta\bar\mu+\cdots\,,\nn
\delta\phi^{RT}&=\frac{\partial \phi^{(0)}}{\partial T}\,\delta T+\frac{\partial \phi^{(0)}}{\partial \bar \mu}\,\delta \bar\mu+\cdots\,.
\end{align}
From this we can read off the near horizon quantities
using the notation of equation  \eqref{eq:nh_exp1}-\eqref{eq:nh_exp3} 
\begin{align}\label{seedqnm}
\delta g^{RT(0)}_{tt}&=\delta g^{RT(0)}_{rr}=\delta g^{RT(0)}_{tr}=-\frac{\delta T}{T},\notag\\
\delta g^{RT(0)}_{ti}&=0,\quad \delta g^{RT(0)}_{ri}=0\,,\quad\delta g^{RT(0)}_{ij}=\frac{\partial g^{(0)}_{ij}}{\partial T}\,\delta T+\frac{\partial g^{(0)}_{ij}}{\partial \bar\mu}\,\delta \bar\mu\,,\nn
\delta a_{t}^{RT(0)}&=-\delta\bar \mu,\quad \delta a_{t}^{RT(1)}=\frac{\partial a_{t}^{(0)}}{\partial T}\,\delta T+\frac{\partial a_{t}^{(0)}}{\partial \bar\mu}\,\delta \bar\mu\,,\notag\\
\delta a_{r}^{RT(0)}&=-\delta\bar \mu,\quad \delta a_{r}^{RT(1)}=\frac{\delta T}{T}\,a_{t}^{(0)} - 4\pi T g^{(1)}\delta \bar \mu\,,\nn
\delta a_{i}^{RT(0)}&=0\,,\qquad
\delta\phi^{RT(0)}=\frac{\partial \phi^{(0)}}{\partial T}\,\delta T+\frac{\partial \phi^{(0)}}{\partial \bar \mu}\,\delta \bar\mu\,.
\end{align}
Notice that this perturbation has $v^i=0$, $w=-\delta\bar \mu$ and $p=4\pi\delta T$ which clearly solves 
\eqref{hconstrainta}-\eqref{hiconstraint2b} for vanishing frequency, $\omega=0$.

We now introduce a small parameter $\varepsilon$ which will be used to perturbatively construct a real time diffusive mode. Following
\cite{Donos:2017gej}, and recalling \eqref{peeeq}, the perturbation is taken to be of the form
\begin{align}\label{pert1}
\delta g_{\mu\nu}&=e^{-i\omega [t+S(r)]+i\varepsilon k_{i}x^{i}}\,\left(\delta g^{RT}_{\mu\nu}+\varepsilon\,\delta g_{[1]\mu\nu}+\varepsilon^{2}\,\delta g_{[2]\mu\nu}+\cdots\right)\,,\nn
\delta A_{\mu}&=e^{-i\omega [t+S(r)]+i\varepsilon k_{i}x^{i}}\,\left(\delta A^{RT}_{\mu}+\varepsilon\,\delta A_{[1]\mu}+\varepsilon^{2}\,\delta A_{[2]\mu}+\cdots\right)\,,\nn
\delta \phi&=e^{-i\omega [t+S(r)]+i\varepsilon k_{i}x^{i}}\,\left(\delta \phi^{RT}+\varepsilon\,\delta \phi_{[1]}+\varepsilon^{2}\,\delta \phi_{[2]}+\cdots\right)\,,
\end{align}
with the corrections $\delta g_{[m]\mu\nu}$, $\delta A_{[m]\mu}$, $\delta \phi_{[m]}$, $m=1,2,\dots$ being time independent, complex functions
of $(r,x^i)$ that are periodic in the spatial coordinates $x^i$.
We demand that order by order in the expansion in $\varepsilon$, the corrections have near horizon expansions analogous to
\eqref{eq:nh_exp1}-\eqref{eq:nh_exp3}. Specifically,
\begin{align}\label{eq:nh_exp_terms}
\delta g_{[m]tt}&=(4\pi Tr)\,\left(\delta g^{(0)}_{[m]tt}\left(x\right)+{\cal O}(r) \right)\,,\qquad\,
\delta g_{[m]rr}=\frac{1}{(4\pi Tr)}\,\left( \delta g_{[m]rr}^{(0)}\left(x\right)+{\cal O}(r)\right),\nn
\delta g_{[m]ij}&=\delta g_{[m]ij}^{(0)}\left(x\right)+{\cal O}(r),
\qquad\qquad\qquad
\delta g_{[m]tr}=\delta g_{[m]tr}^{(0)}\left(x\right)+{\cal O}(r)\,,\nn
\delta g_{[m]tj}&=\delta g_{[m]tj}^{(0)}\left(x\right)+r\delta g_{[m]tj}^{(1)}\left(x\right)+\dots,\quad
\delta g_{[m]rj}=\,\frac{1}{(4\pi Tr)}\,\left(\delta g_{[m]rj}^{(0)}\left(x\right)+r\delta g_{[m]rj}^{(1)}\left(x\right)+\dots  \right)\,,\nn
\delta A_{[m]t}&=\delta a_{[m]t}^{(0)}(x)+r\,\delta a_{{[m]}t}^{(1)}(x)+\cdots\,,\quad
\delta A_{[m]r}=\frac{1}{(4\pi Tr)}\left( \delta a_{[m]r}^{(0)}(x)+r\,\delta a_{{[m]}r}^{(1)}(x)+\cdots\right)\,,\nn
\delta A_{[m]j}&=\delta a_{[m]j}^{(0)}(x)+{\cal O}(r)\,,\quad
\delta \phi_{[m]}=\delta\phi^{(0)}_{[m]}(x)+{\cal O}(r)\,,
\end{align}
with the analogue of the conditions in \eqref{eq:nh_exp3} satisfied for each $[m]$.

\subsection{Dispersion relations for the diffusion modes}
We now explain how we can obtain the dispersion relations for the perturbation, order by order as an expansion in $\varepsilon$.
We will first show how solving the constraint equations \eqref{hconstrainta}-\eqref{hiconstraint2b} on the horizon, to a certain order
in $\varepsilon$, is enough to obtain the leading order dispersion relation for $\omega$ as a function of the wave vector $k_{i}$ in
terms of the horizon DC conductivity, obtained from a Stokes flow on the horizon, and the thermodynamic susceptibilities. 
As already noted, the arguments in appendix \ref{contex} ensure that the full perturbation will solve all the equations of motion.
Some additional subtleties are discussed in appendix \ref{app:expansion}.

From \eqref{pert1},\eqref{eq:nh_exp_terms} and the analogue of \eqref{eq:nh_exp3}, 
the expansion at the horizon that we consider takes the form
\begin{align}\label{eq:hor_fluid_exp}
\omega&=\varepsilon\,\omega_{[1]}+\varepsilon^{2}\,\omega_{[2]}+\cdots\,,\nn
p&=e^{i \varepsilon k_{i}x^{i}}\,\left(4\pi \delta T+\varepsilon\, p_{[1]}+\varepsilon^2\, p_{[2]}  +\cdots\right)\,,\nn
v_{i}&=e^{i \varepsilon k_{i}x^{i}}\left(\varepsilon\,v_{[1]i}+\varepsilon^2\,v_{[2]i}\cdots\right)\,,\nn
 w&=e^{i \varepsilon k_{i}x^{i}}\,\left(-\delta\bar \mu+\varepsilon\,w_{[1]} +\varepsilon^2\,w_{[2]}+\cdots\right)\,,\nn
\delta g_{ij}^{(0)}&=e^{i \varepsilon k_{i}x^{i}}\,\left(\frac{\partial g^{(0)}_{ij}}{\partial T}\,\delta T+\frac{\partial g^{(0)}_{ij}}{\partial \bar\mu}\,\delta \bar\mu +\varepsilon\, \delta g_{[1]ij}^{(0)}+\cdots\right)\,,\nn
\delta\phi^{(0)}&=e^{i \varepsilon k_{i}x^{i}}\,\left(\frac{\partial \phi^{(0)}}{\partial T}\,\delta T+\frac{\partial \phi^{(0)}}{\partial \bar\mu}\,\delta \bar \mu +\varepsilon\,\delta\phi_{[1]}^{(0)}+\cdots \right)\,,
\end{align}
where $v_{[m]i}\equiv -\delta g_{[m]ti}^{(0)}$, $w_{[m]}\equiv \delta a_{[m]t}^{(0)}$ and $p_{[m]}\equiv -4\pi T\delta g_{[m]rt}^{(0)}$ (see \eqref{eq:nh_exp_terms}).
At leading order in $\varepsilon$, the scalar constraint equations
\eqref{hconstrainta} and \eqref{hconstrainttwo} read 
\begin{align}
&\nabla_{i}v_{[1]}^{i}=\frac{i\omega_{[1]}}{2}\left(\delta T \,g_{(0)}^{ij} \frac{\partial g^{(0)}_{ij}}{\partial T} +\delta\bar\mu \,g_{(0)}^{ij} \frac{\partial g^{(0)}_{ij}}{\partial \bar \mu}\right)\,,\nn
&\nabla_j\left( Z^{(0)} \left(-i\,k^{j}\delta\bar\mu+ \nabla^{j}w_{[1]} +v_{[1]}^{j}a_{t}^{(0)}\right) \right) =\nn
&\qquad\qquad i\omega_{[1]} \,\left(\frac{1}{2}Z^{(0)}a^{(0)}_{t}g_{(0)}^{ij}  \frac{\partial g^{(0)}_{ij}}{\partial T}+\partial_{\phi} Z^{(0)}a^{(0)}_{t}\, \frac{\partial \phi^{(0)}}{\partial T}+Z^{(0)} \frac{\partial a_{t}^{(0)}}{\partial T}\right)\delta T\nn
&\qquad\qquad +i\omega_{[1]} \,\left(\frac{1}{2}Z^{(0)}a^{(0)}_{t}g_{(0)}^{ij} \frac{\partial g^{(0)}_{ij}}{\partial \bar\mu}+\partial_{\phi} Z^{(0)}a^{(0)}_{t}\,\frac{\partial \phi^{(0)}}{\partial \bar\mu}+Z^{(0)}\frac{\partial a_{t}^{(0)}}{\partial\bar\mu}\right)\delta \bar \mu\,.
\end{align}
while the vector constraint equation \eqref{hiconstraint2b} has the form
\begin{align}
&-2\,\nabla^{j}\nabla_{(j}{v_{[1]i)}}+\nabla_{i}p_{[1]} -Z^{(0)}a^{(0)}_{t}\,\nabla_{i}w_{[1]} +\nabla_{i}\phi^{(0)} 
v_{[1]}^{j} \nabla_{j}\phi^{(0)}
+ik_{i}4\pi\,\delta T+Z^{(0)}a^{(0)}_{t}\,ik_{i}\delta\bar \mu
\nn
&\qquad -i\,\omega_{[1]}\,\nabla_{i}\phi^{(0)}\,\left(\frac{\partial \phi^{(0)}}{\partial T}\,\delta T+ \frac{\partial \phi^{(0)}}{\partial \bar{\mu}}\,\delta \bar{\mu} \right)+i\,\omega_{[1]}\,\nabla^{k}\left(\frac{\partial g^{(0)}_{ki}}{\partial T}\,\delta T+ \frac{\partial g^{(0)}_{ki}}{\partial \bar{\mu}}\,\delta \bar{\mu}\right)=0
\end{align}
At this point we pause to comment on the structure of this system of equations, which will have echoes at higher orders. 
Specifically, they are a sourced version of the horizon Stokes flow equations which were identified in \cite{Donos:2015gia,Banks:2015wha}
to calculate the DC conductivity (see \eqref{DCone},\eqref{eq:J_hor2ap}). In particular, 
 the temperature gradient and electric field in \eqref{DCone},\eqref{eq:J_hor2ap} are given by $\bar\zeta_{i}=-ik_{i}\delta T /T$ and $\bar E_{i}=-ik_{i}\delta\bar \mu$ and the source terms are parametrised by $\omega_{[1]}$.
Following the arguments of \cite{Donos:2015gia,Banks:2015wha}, as long as the horizon does not have any Killing vectors, the unknown variables $w_{[1]}$, $p_{[1]}$ and $v_{[1]}^{i}$ are fixed up to global shifts of the horizon scalars $w_{[1]}$ and $p_{[1]}$ by constants which we call $\delta\bar{\mu}_{[1]}$ and $4\pi\,\delta T_{[1]}$. We therefore see that it is not possible at this order in perturbation theory to fix these horizon zero modes for the functions $w_{[1]}$ and $p_{[1]}$. However, imposing periodic boundary conditions puts strong constraints on the sources of these equations which appear on the right hand side. On one hand, it will be one of the significant ingredients in fixing the frequency $\omega$ order by order. On the other hand, as we will see in appendix \ref{app:expansion}, the constants $\delta\bar{\mu}_{[1]}$ and $4\pi\,\delta T_{[1]}$ will be fixed by demanding existence of $w_{[3]}$ and $p_{[3]}$ i.e. at third order in perturbation theory. This is the structure one encounters at each order in the $\varepsilon$ expansion. For bookkeeping, we will subtract the zero modes according to
\begin{align}\label{eq:hmode_split}
w_{[i]}=\hat{w}_{[i]}+\delta\bar{\mu}_{[i]}, \qquad p_{[i]}=\hat{p}_{[i]}+4\pi\,\delta T_{[i]}
\end{align}
with the hatted variables having zero average over a period and are therefore uniquely fixed after solving the system of constraints.

To proceed, we multiply by $\sqrt{g_{(0)}}$, and integrate the above equations over a spatial period. Using the definitions of the thermodynamic susceptibilities given in \eqref{susdef} we obtain two conditions which can be written in matrix form as
\begin{align}\label{dcmat}
i\omega_{[1]}\left( \begin{array}{cc}
T^{-1}c_{\mu} &\xi\\
\xi &\chi
\end{array} \right)
\left( \begin{array}{c}
\delta T\\
\delta\bar \mu
\end{array} \right)=0\,.
\end{align}
Assuming that the matrix of susceptibilities is an invertible matrix, which is generically the case, we deduce that $\omega_{[1]}=0$. 
For later reference, notice that we do not assume here that the matrix is positive definite as required for thermodynamical stability.
Thus, at this order in $\varepsilon$ the full set of constraints \eqref{hconstrainta}-\eqref{hiconstraint2b} reduce to 
\begin{align}\label{ordderonece}
&\nabla_{i}v_{[1]}^{i}=0\,,\nn
&\nabla_j\left( Z^{(0)} \left(-i\,k^{j}\delta\bar \mu+ \nabla^{j}w_{[1]} +v_{[1]}^{j}a_{t}^{(0)}\right) \right) =0\,,\nn
&-2\,\nabla^{j}\nabla_{(j}v_{[1]i)}+\nabla_{i}p_{[1]} -Z^{(0)}a^{(0)}_{t}\,\nabla_{i}w_{[1]} +v_{[1]}^{j} \nabla_{j}\phi^{(0)}\nabla_{i}\phi^{(0)} +ik_{i}4\pi\,\delta T+Z^{(0)}a^{(0)}_{t}\,ik_{i}\delta\bar \mu=0\,,
\end{align}
which are now precisely the Stokes equations that are used in determining the DC conductivity given in \eqref{DCone},\eqref{eq:J_hor2ap} 
provided that, as above, we identify the sources $\bar\zeta_{i}=-ik_{i}\delta T /T$ and $\bar E_{i}=-ik_{i}\delta\bar \mu$. These can be uniquely solved for
$\hat w_{[1]}$, $\hat p_{[1]}$ and $v_{[1]}$ (provided that the horizon does not have Killing vectors).
Thus, after integrating over the horizon we can deduce, in particular, that 
\begin{align}
4\pi T i \oint \sqrt{g_{(0)}}v_{[1]}^{i}=T\bar\alpha^{ij}_Hk_j\delta\bar \mu+\bar\kappa^{ij}_Hk_j\delta T\,,\nn
i\oint \sqrt{g_{(0)}}Z^{(0)} \left(-i\,k^{j}\delta\bar \mu+ \nabla^{j}w_{[1]} +v_{[1]}^{j}a_{t}^{(0)}\right) =\sigma^{ij}_Hk_j\delta\bar \mu+\alpha^{ij}_Hk_j\delta T\,,
\end{align}
where $\sigma^{ij}_H$, $\alpha^{ij}_H$, $\bar\alpha^{ij}_H$, $\bar\kappa^{ij}_H$ are the sub-matrices of the full thermoelectric horizon DC conductivity (see \eqref{bigform2}). When the DC conductivities of the dual field theory are finite, as in the case of explicit breaking of translations,
then these are in fact the same as the DC conductivity of the dual field theory, which we will denote
by $\sigma^{ij}$, $\alpha^{ij}$, $\bar\alpha^{ij}$, $\bar\kappa^{ij}$, respectively. 
For the time-reversal invariant backgrounds we are considering it will be useful to recall that $\sigma$ and $\bar \kappa$ are symmetric matrices, while $\alpha=\bar\alpha^T$.

We now examine the constraint equations at second order in $\varepsilon$. 
The scalar constraints \eqref{hconstrainta} and \eqref{hconstrainttwo} give
\begin{align}\label{secorderce}
&\nabla_{i}v_{[2]}^{i}=\frac{i\omega_{[2]}}{2}\left(\delta T \,g_{(0)}^{ij} \frac{\partial g^{(0)}_{ij}}{\partial T} +\delta \bar \mu \,g_{(0)}^{ij} \frac{\partial g^{(0)}_{ij}}{\partial \bar \mu}\right)-ik_{i}v_{[1]}^{i}\,,\nn
&\nabla_j\left( Z^{(0)} \left(i\,k^{j}\delta\bar{\mu}_{[1]}+ \nabla^{j}\hat{w}_{[2]} +g_{(0)}^{ji}v_{[2]j}a_{t}^{(0)}\right) \right)=
-ik_{j}\left( Z^{(0)} \left(-i\,k^{j}\delta\bar \mu+ \nabla^{j}\hat{w}_{[1]} +g_{(0)}^{ji}v_{[1]j}a_{t}^{(0)}\right) \right) \nn
&\qquad\qquad-i\nabla_j\left( Z^{(0)} k^{j}\hat{w}_{[1]} \right) +i\omega_{[2]} \,\left(\frac{1}{2}Z^{(0)}a^{(0)}_{t}g_{(0)}^{ij}  \frac{\partial g^{(0)}_{ij}}{\partial T}+\partial_{\phi} Z^{(0)}a^{(0)}_{t}\, \frac{\partial \phi^{(0)}}{\partial T}+Z^{(0)} \frac{\partial a_{t}^{(0)}}{\partial T}\right)\delta T\nn
&\qquad\qquad +i\omega_{[2]} \,\left(\frac{1}{2}Z^{(0)}a^{(0)}_{t}g_{(0)}^{ij} \frac{\partial g^{(0)}_{ij}}{\partial \bar \mu}+\partial_{\phi} Z^{(0)}a^{(0)}_{t}\,\frac{\partial \phi^{(0)}}{\partial \bar \mu}+Z^{(0)}\frac{\partial a_{t}^{(0)}}{\partial\bar \mu}\right)\delta\bar  \mu\,.
\end{align}
and we will not explicitly write down the vector equation \eqref{hiconstraint2b}.
We remind the reader that the constants $\delta\bar{\mu}_{[1]}$ and $\delta T_{[1]}$ have not been fixed yet and will be fixed by demanding existence of the solution at third order in the perturbative expansion, as discussed further in appendix \ref{app:expansion}. 
Furthermore, the zero modes $\delta\bar{\mu}_{[2]}$ and $4\pi\,\delta T_{[2]}$ of the 
 functions $w_{[2]}$ and $p_{[2]}$ that we extracted according to \eqref{eq:hmode_split} will only be fixed by demanding existence of the 
 solution at fourth order in the $\varepsilon$ expansion.

Even without knowing the value of the constants $\delta\bar{\mu}_{[1]}$ and $\delta T_{[1]}$, existence of the solution at second order will constrain the frequency $\omega_{[2]}$ and the constants $\delta\bar{\mu}$ and $\delta T$. Indeed, integrating the two equations \eqref{secorderce} over a period we find that 
\begin{align}\label{modeseqn}
i\omega_{[2]}\left(T^{-1}c_{\mu}\delta T+\xi\,\delta\bar \mu \right)-T^{-1}\bar{\kappa}^{ij}k_{i}k_{j}\,\delta T-\alpha^{ij}k_{i}k_{j}\,\delta\bar\mu=&0\,,\nn
i\omega_{[2]}\left(\xi\,\delta T+\chi\,\delta\bar \mu \right)-\alpha^{ij}k_{i}k_{j}\,\delta T-\sigma^{ij}k_{i}k_{j}\,\delta\bar \mu=&0\,,
\end{align}
where we used the fact that $\alpha^{ij}k_{i}k_{j}=\bar \alpha^{ij}k_{i}k_{j}$, for the backgrounds we are considering.

The algebraic system \eqref{modeseqn} is exactly the same as that considered in \cite{Donos:2017gej} (see also \cite{Hartnoll:2014lpa}) and we can immediately obtain the 
two eigenfrequencies $i\omega_{[2]}^{\pm}$ associated with the diffusion modes.
Defining
\begin{align}
\bar{\kappa}(k)\equiv\bar{\kappa}^{ij}k_{i}k_{j},\quad \alpha(k)\equiv\alpha^{ij}k_{i}k_{j},\quad \sigma(k)\equiv \sigma^{ij}k_{i}k_{j}\,,
\end{align}
we obtain the generalised Einstein relation
\begin{align}\label{twoevals}
i\omega_{[2]}^{+}\, i\omega_{[2]}^{-}&=\frac{\kappa(k)}{c_\rho}\frac{\sigma(k)}{\chi}\,,\nn
i\omega_{[2]}^{+}+ i\omega_{[2]}^{-}&=\frac{\kappa(k)}{c_{\rho}}+\frac{\sigma(k)}{\chi}+\frac{T\,\left[\chi\,\alpha(k)-\xi\,\sigma(k) \right]^{2}}{c_{\rho}\chi^{2}\sigma(k)}\,,
\end{align}
where $c_{\rho}= c_{\mu}-\frac{T\xi^{2}}{\chi}$ was given in \eqref{ceerho} and we have defined
\begin{align}
\kappa(k)\equiv \bar{\kappa}(k)-\frac{\alpha^{2}(k)T}{\sigma(k)}\,.
\end{align}
This is the universal result concerning the dispersion relations for the diffusive modes associated with the conserved heat and electric currents for holographic lattices.

\subsection{Comments}\label{commentssec}
Recall that $\bar \kappa^{ij}$ is the thermal DC conductivity for zero applied electric field. 
On the other hand the thermal DC conductivity for zero electric current, $\kappa^{ij}$, is given by $\kappa^{ij}=\bar\kappa^{ij}-T(\bar \alpha\sigma^{-1}\alpha)^{ij}$.
Despite the notation, note that, in general, $\kappa(k)\ne \kappa^{ij}k_i k_j$. 

The simplest dispersion relations occur for charge neutral background black holes with vanishing gauge fields.
In this case we have $\alpha^{ij}=\xi=0$ and hence $\bar\kappa^{ij}=\kappa^{ij}$ leading to the
simple Einstein relations\footnote{In a charge neutral holographic setting, with translations explicitly broken using scalar fields as in \cite{Andrade:2013gsa},
the first diffusive mode in \eqref{disqz2} was numerically constructed in \cite{Davison:2014lua}.}
\begin{align}\label{disqz2}
i\omega=\varepsilon^2\frac{\kappa^{ij}k_ik_j}{c_\mu}+\dots\,,\qquad
i\omega=\varepsilon^2\frac{\sigma^{ij}k_ik_j}{\chi}+\dots\,.
\end{align}
In the special case of translationally invariant black holes with vanishing gauge fields\footnote{This is the setting where the holographic Einstein relation for electric charge diffusion was first discussed in \cite{Iqbal:2008by}.},
the electric DC conductivity of the dual field theory is still finite and there is a corresponding charge diffusive mode as in \eqref{disqz2}.
On the other hand the thermal DC conductivity is infinite and there is no thermal diffusive mode.
 
To make some additional comments, we first recall some results concerning the DC conductivity for 
perturbative holographic lattices for which translations are broken weakly. Such lattices
have a black hole horizon that is a perturbation about a flat horizon, parametrised by a small number $\lambda$.
In \cite{Donos:2015gia,Banks:2015wha} it was shown that the DC conductivity has the expansion
\begin{align}\label{phlatres}
&\bar\kappa^{ij}=(L^{-1})^{ij}4\pi sT+\mathcal{O}(\lambda^{-1})\,,\, \alpha^{ij}=\bar\alpha^{ij}=(L^{-1})^{ij}4\pi \rho+\mathcal{O}(\lambda^{-1})\,,\nn
&\sigma^{ij}=(L^{-1})^{ij}\frac{4\pi \rho^2}{s}+\mathcal{O}(\lambda^{-1})\,.
\end{align}
where the matrix $L_{ij}$ is proportional to $\lambda^{2}$ and depends on the leading order deviations of the horizon from the translationally
invariant configuration. An explicit expression for $L$ in terms of the spatially modulated horizon was given in \cite{Donos:2015gia,Banks:2015wha}. 
It was also shown in \cite{Donos:2015gia,Banks:2015wha} that both the thermal DC conductivity at zero current flow, $\kappa^{ij}$, and the electric conductivity at zero heat current flow, $\sigma_{Q=0}^{ij}$, appear at a higher order in the expansion. Explicitly,
when $\rho\ne0$ we have
\begin{align}\label{phlatres2}
\sigma^{ij}_{Q=0}=\left.\frac{1}{4\pi}s Z^{(0)}g^{ij}_{(0)}\right|_{\lambda=0}+\mathcal{O}(\lambda)\,,\qquad
T\kappa^{ij}=\left.\frac{s^3 T^2Z^{(0)}g^{ij}_{(0)}}{4\pi\rho^2}\right|_{\lambda=0}+\mathcal{O}(\lambda)\,.
\end{align}
Notice, in particular, $T\kappa^{ij}=\frac{s^2T^2}{\rho^2}\sigma^{ij}_{Q=0}+\mathcal{O}(\lambda)$. When $\rho=0$ the
expression for $\sigma^{ij}_{Q=0}$ is still valid but we can no longer calculate $T\kappa^{ij}$ perturbatively
as the leading order piece is infinite.

Using the results \eqref{phlatres} in \eqref{twoevals} we
obtain the dispersion relations of the two diffusive modes:
\begin{align}\label{bloppy}
i\omega^+&=\frac{\rho^2\chi}{c_\rho\rho^2\chi +T(\xi\rho-s\chi)^2}\kappa^{ij}k_ik_j+\mathcal{O}(\lambda)\,,\nn
&=\frac{(sT)^2}{(sT)^2\chi-2(sT\rho) T\xi +\rho^2 Tc_\mu}\sigma^{ij}_{Q=0}k_ik_j+\mathcal{O}(\lambda)\,,\nn
i\omega^-&=\frac{4\pi(c_\rho\rho^2\chi+T(\xi\rho-s\chi)^2)}{c_\rho s\chi^2}(L^{-1})^{ij}k_ik_j+\mathcal{O}(\lambda^{-1})\,.
\end{align}
In particular, the first diffusive mode is of order $\lambda^0$, but the second is of order $\lambda^{-2}$ and
hence appears parametrically further down the imaginary axis (while, of course, still going to the origin when $k_i\to 0$). 
If we now consider the translationally invariant case by taking $\lambda\to 0$,
we find that we have one diffusive mode with a dispersion relation that satisfies an Einstein relation in terms of the finite DC conductivity $\kappa^{ij}$, or $\sigma^{ij}_{Q=0}$ (which is also valid when $\rho=0$), as given in the first two lines of \eqref{bloppy}. Such a diffusive mode was discussed in the context of hydrodynamics in
\cite{Davison:2015taa}.

We can also consider spontaneous breaking of translations
with the addition of a small explicit breaking, with dimensionless strength $\lambda$,
as recently discussed for specific helical lattices in \cite{Andrade:2017cnc}. 
For temperatures just below $T_c$ one can develop a double expansion in both
$\lambda$ and $(1-T/T_c)^{1/2}$, with the exponent in the latter the expected behaviour for standard mean field phase transitions.
The matrix $L_{ij}$ in \eqref{phlatres} will then also have have a double expansion.
For the case that $\lambda >> (1-T/T_c)^{1/2}$ we can expand, schematically,
$L^{-1}\sim \lambda^{-2}[a_1+\dots]$, where $a_1$ is a horizon quantity that can be calculated as in 
\cite{Donos:2015gia,Banks:2015wha} and the neglected terms are a double expansion in $(1-T/T_c)^{1/2}/\lambda$ and $\lambda$.
In particular, in this limit we see that the DC conductivity is dominated by the explicit breaking terms, as expected.
Similarly, for $\lambda << (1-T/T_c)^{1/2}$ we have
$L^{-1}\sim (1-T/T_c)^{-1}[a_2+\dots ]$, where
the neglected terms are an expansion in $\lambda/(1-T/T_c)^{1/2}$ and $(1-T/T_c)^{1/2}$.
This result
explains a feature of the DC conductivity that was 
found numerically in figure 9 of \cite{Andrade:2017cnc}. 
It is also worth noting that in the
case that $\lambda << (1-T/T_c)^{1/2}$ this drop in the DC conductivity, combined with sum rules,
implies that in the AC conductivity the spectral weight will move from the Drude peak to mid frequencies, as seen in
the example of \cite{Andrade:2017cnc}.
For the case of spontaneous breaking
with a small explicit breaking there will, of course, still be two diffusive modes with dispersion relations as
in \eqref{bloppy}, and both can be expanded in terms of $\lambda$ and $(1-T/T_c)^{1/2}$.
Note that in \cite{Andrade:2017cnc}, for a specific setting of pinned helical phases, only the first diffusive mode in \eqref{bloppy} was discussed.

Our final comment concerns instabilities of the background black hole solutions. In particular,
the dispersion relations for the diffusive modes given in
\eqref{twoevals}
allow us to make sharp statements concerning the relation between thermodynamic instability
and dynamical instability of the holographic lattice black hole solutions. 
In the simplest case, when the gauge field is zero we know that when
$c_\mu$ or $\chi$ is negative then we have a thermodynamic instability. But from \eqref{disqz2} we immediately deduce that there is a 
quasinormal mode with a pole in the upper half of the complex frequency plane and this leads to a dynamical instability of the black hole solution. 

Turning now to general black hole backgrounds with non-vanishing gauge-field, we can write the equation for the diffusive modes in \eqref{modeseqn} as
\begin{align}\label{dcmat2}
\left[i\omega_{[2]}\left( \begin{array}{cc}
1 &0\\
0 &1
\end{array} \right)-
\left( \begin{array}{cc}
\chi &T\xi\\
T\xi &Tc_\mu
\end{array} \right)^{-1}
\left( \begin{array}{cc}
\sigma(k) &T\alpha(k)\\
T\alpha(k) &T\bar{\kappa}(k)
\end{array} \right)\right]
\left( \begin{array}{c}
\delta\bar\mu\\
\delta T/T
\end{array} \right)=0\,.
\end{align}
Since the thermoelectric horizon DC conductivity is a positive definite matrix, we see that a negative eigenvalue in the susceptibility matrix appearing in \eqref{dcmat2}, associated with a thermodynamic instability, will again give rise to a quasi-normal mode
in the upper half complex frequency plane and, correspondingly, a dynamical instability for the black hole.

\section{Final comments}
In this paper we showed how the quasi-normal modes associated with heat and charge diffusion can be constructed for holographic lattices in a long-wavelength,
perturbative expansion.
In particular the construction allowed us to derive the dispersion relation of the diffusive modes
in terms of horizon DC conductivities, obtained from solutions to a Stokes flow on the horizon, and static susceptibilities. 
This constitutes a derivation of a generalised Einstein relation.

We considered a class of gravitational theories with a specific matter content, but it is clear that the main results should apply to more general theories, including the possibility of having more gauge fields in the bulk and hence additional conserved charges in the dual field theory. We focussed on studying static geometries for simplicity, but it should be possible
to relax this condition utilising the holographic understanding of transport currents presented in \cite{Donos:2015bxe,Donos:2017oym}. 
Similarly, the extension
to higher derivative theories of gravity should also be possible using the results in \cite{Donos:2017oym}.

The derivation of the dispersion relations started with the identification of the quasinormal mode at $\omega=k_i=0$, namely \eqref{seedqnm}.
This was possible because this diffusion mode is associated with conserved quantities. This was then used to perturbatively
construct the quasinormal modes in a neighbourhood of $\omega=k_i=0$. In particular, the analysis of the constraint equations on the stretched horizon was sufficient to obtain the dispersion relation for the quasinormal mode. It is clear that this procedure will work for the quasinormal modes
associated with any conserved quantity\footnote{It also seems likely that if one is given a specific quasinormal mode for some $(\omega_0,k_0)\ne0$ then it should also be possible
to obtain the dispersion relation for the mode for  $(\omega,k)$ close to $(\omega_0, k_0)$. However, in the case, the technical effort required to obtain the specific quasinormal mode 
probably allows one to construct the quasinormal mode for $(\omega,k)\ne (\omega_0, k_0)$ and so it is not clear if this observation is that
significant.}.

There has been some interesting recent discussion of the Goldstone modes that arise from spontaneously broken
translation invariance, with an emphasis on the pinning phenomenon that occurs after adding in a small explicit breaking of translations,
both within holography \cite{Andrade:2017cnc,Alberte:2017cch,Jokela:2017ltu} and from a hydrodynamic point of view 
\cite{Davison:2016hno,Delacretaz:2016ivq,Delacretaz:2017zxd}. In the future, we plan to report on how the methods developed in this paper can be extended to study these modes as well as the Goldstone modes arising in spontaneously broken internal symmetries.

\section*{Acknowledgements}
We would like to thank Tom Griffin for helpful discussions.
The work of JPG is supported by the European Research Council under the European Union's Seventh Framework Programme (FP7/2007-2013), ERC Grant agreement ADG 339140, STFC grant ST/P000762/1, EPSRC grant EP/K034456/1, as a KIAS Scholar and as a Visiting Fellow at the Perimeter Institute. VZ is supported by a Faculty of Science Durham Doctoral Scholarship.

\appendix

\section{Residual gauge invariance}\label{resgi}
The time dependent perturbation that we introduced in section \ref{timedeppert} satisfied ingoing boundary conditions
at the horizon summarised in \eqref{eq:nh_exp1}-\eqref{eq:nh_exp3}. It is illuminating to note that there are some residual gauge and coordinate transformations which near the horizon are given by 
\begin{align}
\delta \Lambda&=e^{-i\omega v_{EF}}\,\left(\delta \lambda^{(0)} (x^{i})+r\,\delta \lambda^{(1)} (x^{i})+,\mathcal{O}(r^{2})\right)\,,\nn
t&\to t +e^{-i\omega v_{EF}} \left( \delta T^{(0)}(x^{j})+\mathcal{O}(r) \right)\,,\nn
r&\to r +e^{-i\omega v_{EF}} \left(r\delta R^{(0)}(x^{j})+\mathcal{O}(r^2)\right)\,,\nn
x^{i}&\to x^{i}+e^{-i\omega v_{EF}}\left(\delta\xi^{(0)}{}^{i}(x^{j})+r\, \delta\xi^{(1)}{}^{i}(x^{j})+\mathcal{O}(r^{2}) \right)\,.
\end{align}
These are consistent with  \eqref{eq:nh_exp1}, \eqref{eq:nh_exp2} and induce the following
transformations 
\begin{align}\label{gt1}
\delta g^{(0)}_{tt}&\to \delta g^{(0)}_{tt}+2i\omega \delta T^{(0)}-\delta R^{(0)},\quad
\delta g^{(0)}_{rr}\to \delta g_{rr}^{(0)}+(1-\frac{2i\omega}{4\pi T})\delta R^{(0)},\nn
\delta g^{(0)}_{ij}&\to \delta g_{ij}^{(0)}+2\,\nabla_{(i}\delta\xi^{(0)}_{j)},
\qquad\quad\,\,\,
\delta g^{(0)}_{tr}\to \delta g_{tr}^{(0)}+i\omega\delta T^{(0)}-\frac{i\omega}{4\pi T}\delta R^{(0)}\,,\nn
\delta g^{(0)}_{ti}&\to \delta g_{ti}^{(0)}-i\omega\delta\xi^{(0)}_i   \,,
\qquad
\delta g^{(1)}_{ti}\to \delta g_{ti}^{(1)}
-4\pi T\partial_i\delta T^{(0)}-i\omega (g^{(1)}_{ij}\delta\xi^{(0)j}+\delta\xi^{(1)}_i)\,,
\nn
\delta g^{(0)}_{ri}&\to \delta g_{ri}^{(0)}-i\omega\delta\xi^{(0)}_i   \,,
\qquad
\delta g^{(1)}_{ri}\to \delta g_{ri}^{(1)}
+\partial_i\delta R^{(0)}-i\omega (g^{(1)}_{ij}\delta\xi^{(0)j}+\delta\xi^{(1)}_i)+4\pi T\delta\xi^{(1)}_i\,,
\end{align}
as well as
\begin{align}\label{gt2}
\delta a^{(0)}_{t}&\to\delta a^{(0)}_{t}-i\omega\delta\lambda^{(0)},
\qquad
\delta a^{(1)}_{t}\to\delta a^{(1)}_{t}-i\omega\delta\lambda^{(1)}-i\omega a^{(0)}_t\delta T^{(0)}+a^{(0)}_t \delta  R^{(0)}+(\partial_i a^{(0)}_t) \delta \xi ^{(0)i},
\nn
\delta a^{(0)}_{r}&\to\delta a^{(0)}_{r}-i\omega\delta\lambda^{(0)},
\qquad
\delta a^{(1)}_{r}\to\delta a^{(1)}_{r}+(4\pi T-i\omega)\delta\lambda^{(1)}-i\omega a^{(0)}_t\delta T^{(0)}\,,
\nn
\delta a_{i}^{(0)}&\to\delta a_{i}^{(0)}+\partial_i\delta \lambda^{(0)},
\qquad
\delta \phi^{(0)}\to\delta \phi^{(0)}+(\partial_i \phi^{(0)})\delta\xi^{(0)i}\,.
\end{align}
One can check that the constraint equations given in \eqref{hconstrainta}-\eqref{hiconstraint2b} 
are covariant with respect to these transformations.

Notice that the combination $\delta a^{(1)}_{t}-\frac{i \omega}{4\pi T}( \delta a^{(1)}_{t}-\delta a^{(1)}_{r} )$, appearing in the horizon constraint equation \eqref{hconstrainttwo}, 
is invariant under the gauge transformations parametrised by $\delta\lambda^{(1)}$.
Similarly, the combination $\delta g_{ti}^{(1)}
-\frac{i\,\omega}{4\pi T}(\delta g_{ti}^{(1)}-\delta g_{ri}^{(1)})$, appearing
in \eqref{hiconstraint2b}, is invariant under the gauge transformations parametrised by $\delta\xi^{(1)}_i$.
If $i\omega\ne 4\pi T$ we can choose $\delta\lambda^{(1)}$ to set $\delta a^{(1)}_{r}=0$
and $\delta\xi^{(1)}_i$ to set $\delta g_{ri}^{(1)}=0$, but we have not found a need to do this, nor fix any of the above gauge invariances\footnote{We comment that a brief discussion of 
performing  AC and DC calculations in a radial gauge, for Q-lattice constructions,  
appear in section 3 of \cite{Donos:2013eha} and in footnote 10 of \cite{Donos:2014uba}.}.

\section{Evaluating the constraints on the horizon}\label{evalcon}

\subsection{Constraints in the radial decomposition}
We begin by briefly summarising the constraint equations that appear in a Hamiltonian decomposition of the equations of motion using
a radial foliation, following the presentation in appendix A of \cite{Banks:2015wha}. 
We introduce the normal vector $n^\mu$, satisfying $n^\mu n_\mu=1$.
The $D$-dimensional metric $g_{\mu\nu}$ induces a $(D-1)$-dimensional Lorentzian
metric on the slices of constant $r$ via $h_{\mu\nu}=g_{\mu\nu}-n_\mu n_\nu$.
The lapse and shift vectors are given by $n_\mu=N(dr)_\mu$ and $N^\mu=h^\mu{}_\nu r^\mu$, respectively, where
$r^\mu=(\partial_r)^\mu$. The gauge-field components are decomposed via $b_\mu=h_\mu{}^\nu A_\nu$,
$\Phi=-Nn^\mu A_\mu$ and we define $f_{\mu\nu}=\partial_\mu b_\nu-\partial_\nu b_\mu$.

The momenta conjugate to $h_{\mu\nu}$, ${b}_{\mu}$ and $\phi$ are given by
\begin{align}\label{momdefns}
\pi^{\mu\nu}&=-\sqrt{-h}\,\left( K^{\mu\nu}-K\,h^{\mu\nu}\right)\,,\notag\\
\pi^\mu&=\sqrt{-h} Z F^{\mu\rho}n_\rho\,,\nn
\pi_\phi&
=-\sqrt{-h}n^\nu\partial_\nu\phi\,,
\end{align}
respectively, where $K_{\mu\nu}=\frac{1}{2}{\mathcal L}_n h_{\mu\nu}$ is the extrinsic curvature. 
The Hamiltonian, momentum and Gauss law constraints can then be written in the form $H=H^{\nu}=C=0$ where
\begin{align}\label{constraints}
H=&\,-(-h)^{-1/2}\,\left(\pi_{\mu\nu}\pi^{\mu\nu}-\frac{1}{D-2}\,\pi^{2} \right)-\sqrt{-h}\,\left({}^{(D-1)}R-V\right)\nn
&-\frac{1}{2}\,(-h)^{-1/2}\,Z^{-1}h_{\mu\nu}\,\pi^{\mu}\,\pi^{\nu}+\frac{1}{4}\,\sqrt{-h}\,Zf_{\mu\nu}\,f_{\rho\sigma}\,h^{\mu\rho}\,h^{\nu\sigma}\nn
&-\frac{1}{2}(-h)^{-1/2}\pi^2_\phi+\frac{1}{2}\sqrt{-h} h^{\rho\sigma}\partial_\rho\phi\partial_\sigma\phi
\,,\nn
H^{\nu}=&-2\sqrt{-h}\,D_{\mu}\left((-h)^{-1/2}\pi^{\mu\nu}\right)+h^{\nu\sigma}f_{\sigma\rho} \pi^\rho\nn
&-h^{\nu\sigma}b_\sigma\sqrt{-h} D_\rho\left( (-h)^{-1/2}\, \pi^\rho\right)
+h^{\nu\sigma}\partial_\sigma\phi\pi_\phi \,,\nn
C=&\sqrt{-h}\,D_{\mu}\left( (-h)^{-1/2}\,\pi^{\mu}\right)\,,
\end{align}
where $\pi=\pi^{\mu}{}_{\mu}$
and $D_\mu$ is the covariant derivative with respect to $h_{\mu\nu}$. 

The full equations of motion are the above constraint equations combined with the radial equations of motion. The latter consist of 
expressions for $\mathcal{L}_r h_{\mu\nu}$,
$\mathcal{L}_r b_\mu$ and $\mathcal{L}_r\phi$ as well as $\mathcal{L}_r\pi^{\mu\nu}$,$\mathcal{L}_r\pi^{\mu}$,$\mathcal{L}_r\pi_\phi$ and explicit expressions can be found in appendix A of \cite{Banks:2015wha}. This leads to equations that have second order
radial derivatives for $h_{\mu\nu}$, $b_\mu$ and $\phi$.

\subsection{Evaluating constraints for the perturbation}

Consider a general perturbation of the background black hole solution \eqref{eq:DC_ansatz} of the form
$\delta g_{\mu\nu}$, $\delta a_\mu$, $\delta\phi$ with all quantities functions of $(t,r,x^i)$.
On the surfaces of constant $r$ this gives rise to a perturbed normal vector with components given by
\begin{align}
n^{i}&=-(U/F)^{1/2}\,g_{d}^{ij}\,\delta g_{rj},\qquad n^{t}=G^{-1}F^{-1/2}U^{-1/2}\delta g_{tr},\nn
n^{r}&=(U/F)^{1/2}\left( 1-\frac{U}{2F}\,\delta g_{rr}\right)\,.
\end{align}
Furthermore, the corresponding shift and lapse functions are given by
\begin{align}
N^{j}&=g_{d}^{ij}\,\delta g_{ri},\qquad N^{t}=-\frac{1}{GU}\,\delta g_{rt}\,,\nn
N&=(F/U)^{1/2}\left( 1+\frac{1}{2}\,\frac{U}{F}\,\delta g_{rr}\right)\,.
\end{align}
The components of the extrinsic curvature $K_{\mu\nu}$ take the form
\begin{align}\label{genkay}
K^{t}{}_{t}=&\frac{1}{2}\,G^{-1}F^{-1/2}U^{-1/2}\,\left(\partial_{r}\left(GU\right)-\frac{1}{2}\,\frac{U}{F}\,\partial_{r}\left(GU\right) \,\delta g_{rr}\right)+\frac{1}{2}\,G^{-2}F^{-1/2}U^{-3/2}\,\partial_{r}\left(GU\right)\delta g_{tt}\nn
&-\frac{1}{2}\,G^{-1}F^{-1/2}U^{-1/2}\,\left(\partial_{r}\delta g_{tt}+\partial_{j}(GU)\,N^{j}\right)+G^{-1}F^{-1/2}U^{-1/2}\,\partial_{t}\delta g_{tr}\,,\nn
K^{i}{}_{t}=&-\frac{1}{2}\,GF^{-1/2}U^{3/2}\,g_{d}^{ij}\left(-\partial_{r}\left( \frac{1}{GU}\,\delta g_{tj} \right)+\partial_{j}\left(\frac{1}{GU}\,\delta g_{rt} \right) +\frac{1}{GU}\partial_{t}\,\delta g_{rj}  \right)\,,\nn
K^{t}{}_{i}=&\frac{1}{2}\,(U/F)^{1/2}\left(-\frac{1}{GU}\partial_{r}\delta g_{ti}+\frac{\delta g_{tj}}{GU}g_{d}^{kj}\,\partial_{r}g_{dik}+\,\partial_{i}\left(\frac{1}{GU}\,\delta g_{rt} \right)+\frac{1}{GU}\partial_{t}\delta g_{ri} \right)\,,\nn
K^{i}{}_{j}=&\frac{1}{2}\,(U/F)^{1/2}\,\left( g^{ik}_d\partial_{r}g_{dkj}-\frac{U}{2F}g^{ik}_d\partial_{r}g_{dkj}\delta g_{rr}+g^{ik}_d\partial_{r}\delta g_{kj}
-g_d^{il}g_d^{km}\partial_{r}g_{dkj} \delta g_{lm} \right)\nn
&-\frac{1}{2}(U/F)^{1/2}\,\left( \nabla^{i}N_{j}+\nabla_{j}N^{i} \right)\,.
\end{align}

We now turn to the specific perturbation discussed in section \ref{timedeppert}. We want to evaluate the constraints at the horizon
by employing the expansions given in \eqref{eq:nh_exp1}-\eqref{eq:nh_exp3}.
Expanding the extrinsic curvature near the horizon we obtain
\begin{align}\label{kayathor}
K^{t}{}_{t}&\rightarrow e^{-i\omega v_{EF}}\frac{1}{2}\frac{(4\pi T)^{1/2}}{r^{1/2}}\,\left(e^{i\omega v_{EF}}-\frac{1}{2}\delta g^{(0)}_{rr}
-\frac{i \omega}{4\pi T}\delta g^{(0)}_{rr} \right)\,,\nn
K^{i}{}_{t}&\rightarrow e^{-i\omega v_{EF}}\frac{1}{2}\frac{(4\pi T)^{1/2}}{r^{1/2}}\,v^i\,,\nn
K^{t}{}_{i}&\rightarrow e^{-i\omega v_{EF}}\frac{1}{2}\,\frac{1}{(4\pi T)^{1/2}}\frac{1}{r^{1/2}}\,\left(-\delta g_{ti}^{(1)}+\partial_{i}\delta g_{tr}^{(0)}-g^{(1)}_{il}v^{l}+\frac{i\,\omega}{4\pi T}(\delta g_{ti}^{(1)}-\delta g_{ri}^{(1)}) \right)\,,\nn
K^{i}{}_{j}&\rightarrow e^{-i\omega v_{EF}}\frac{1}{2}\frac{1}{(4\pi T)^{1/2}}\frac{1}{r^{1/2}}\,\left( \nabla^{i}v_{j}+\nabla_{j}v^{i} -i\omega\, g_{(0)}^{ik} \delta g^{(0)}_{kj}\right)\,,\nn
K&\rightarrow e^{-i\omega v_{EF}}\frac{1}{2}\,\frac{(4\pi T)^{1/2}}{r^{1/2}}\,\left( e^{i\omega v_{EF}}-\frac{1}{2}\delta g^{(0)}_{rr}+\frac{2}{4\pi T}\,\nabla_{i}v^{i}-\frac{i \omega}{4\pi T}\left(\delta g^{(0)}_{rr}+g_{(0)}^{ij} \delta g^{(0)}_{ij} \right)\right)\,.
\end{align}
In the above we have only kept background terms at leading order $\mathcal{O}(r^{-1/2})$ since these are the only ones that will contribute in our calculation. Furthermore, the covariant derivative is with respect to the metric $g^{(0)}_{ij}$, which is also used to raise and lower indices.

We next consider the following quantity which appears in the momentum constraint 
\begin{align}
W_{\nu}=&D_{\mu}\left((-h)^{-1/2}\pi^{\mu}{}_{\nu}\right)=-D_{\mu}K^{\mu}{}_{\nu}+D_{\nu}K\,,\nn
=&-(-h)^{-1/2}\,\partial_{\mu}\left(\sqrt{-h}\,K^{\mu}{}_{\nu}\right)+\frac{1}{2}\,\partial_{\nu}h_{\kappa\lambda}K^{\kappa\lambda}+\partial_{\nu}K\,.
\end{align}
Expanding at the horizon we find the following individual components 
\begin{align}\label{wstuf}
W_{t}&\rightarrow e^{-i\omega v_{EF}} \frac{(4\pi T)^{1/2}}{r^{1/2}}\,\left(-\frac{1}{2}\nabla_{i}v^{i}+\frac{i\omega}{4}g_{(0)}^{ij}\delta g^{(0)}_{ij}-\frac{i\omega}{8\pi T}\left(2\,\nabla_{i}v^{i}-i\omega g_{(0)}^{ij}\delta g^{(0)}_{ij} \right)\right)\,,\nn
W_{i}&\rightarrow \frac{1}{2}e^{-i\omega v_{EF}}\frac{1}{(4\pi T)^{1/2}}\frac{1}{r^{1/2}}\,\Bigl[-2\,\nabla^{j}\nabla_{(j}v_{i)} +\nabla_{i}p^{\prime} \nn
&+i\omega\left(-\delta g_{ti}^{(1)}+\partial_{i}\delta g_{tr}^{(0)}-g^{(1)}_{il}v^{l}+\frac{i\,\omega}{4\pi T}(\delta g_{ti}^{(1)}-\delta g_{ri}^{(1)})+\nabla^{k}\delta g^{(0)}_{ki} \right) \Bigr]\,,
\end{align}
with
\begin{align}
p^{\prime}=-2\pi T(\delta g_{tt}^{(0)}+\delta g_{rr}^{(0)})+2\,\nabla_{j}v^{j}+i \omega\left(\delta g^{(0)}_{tt}-2\delta g^{(0)}_{tr}-g_{(0)}^{ij} \delta g^{(0)}_{ij} \right)\,.
\end{align}

Another quantity that enters the constraints is the momentum of the scalar field. At leading order in $r$ we have
\begin{align}
\pi_{\phi}&\rightarrow -\sqrt{g_{(0)}}\,e^{-i\omega v_{EF}}\left( v^{i} \partial_{i}\phi^{(0)}-i\omega \delta\phi^{(0)}\right)\,.
\end{align}
We next turn to the gauge field. We find 
\begin{align}\label{fathor}
F_{tr}&\rightarrow -a^{(0)}_{t}+e^{-i\omega v_{EF}}\,\left( -\delta a^{(1)}_{t}+\frac{i \omega}{4\pi T}\left( \delta a^{(1)}_{t}-\delta a^{(1)}_{r} \right) \right)\,,\nn
F_{ir}&\rightarrow e^{-i\omega v_{EF}}\frac{1}{4\pi T r}\,\left( \partial_{i}w +i\omega \delta a^{(0)}_{i} \right)
+e^{-i\omega v_{EF}}\frac{1}{4\pi T}\left( \partial_{i}\delta a_{r}^{(1)}-4\pi T\delta a^{(1)}_{i}
+i\omega \delta a^{(1)}_{i}  \right),\nn
F_{ti}&\rightarrow -e^{-i\omega v_{EF}}\,\left( \partial_{i}w+i\omega\delta a_{i}^{(0)} \right) - r\left( \partial_{i}a_{t}^{(0)}+e^{-i\omega v_{EF}}\partial_{i}\delta a_{t}^{(1)}+i\omega\,e^{-i\omega v_{EF}}\delta a_{i}^{(1)} \right)\,,\nn
F_{ij}&\rightarrow 2e^{-i\omega v_{EF}}\,\partial_{[i}\delta a_{j]}^{(0)}\,,\nn
F^{tr}&\rightarrow a^{(0)}_{t}+e^{-i\omega v_{EF}}\,\left( \delta a^{(1)}_{t}-\frac{i \omega}{4\pi T}\left( \delta a^{(1)}_{t}-\delta a^{(1)}_{r}  \right) + \left( \delta g_{tt}^{(0)}-\delta g_{rr}^{(0)} \right) a^{(0)}_{t} + \frac{1}{4\pi T}g_{(0)}^{ij}v_j \partial_{i}a_{t}^{(0)}\right)\,,\nn
F^{ir}&\rightarrow e^{-i\omega v_{EF}}\,g_{(0)}^{ij}\left( \partial_{j}w +i\omega \delta a^{(0)}_{j} +v_{j}a_{t}^{(0)}\right)\,.
\end{align}
and thus the associated momentum has the expansion
\begin{align}\label{pimuathor}
\pi^{i}&\rightarrow \sqrt{g_{(0)}}Z^{(0)} e^{-i\omega v_{EF}}\,g_{(0)}^{ij}\left( \partial_{j}w +i\omega \delta a^{(0)}_{j} +v_{j}a_{t}^{(0)}\right)\,,\nn
\pi^{t}&\rightarrow \sqrt{g_{(0)}}Z^{(0)}e^{-i\omega v_{EF}}\,\left( \delta a^{(1)}_{t}-\frac{i \omega}{4\pi T}\left( \delta a^{(1)}_{t}-\delta a^{(1)}_{r} \right) + \frac{1}{4\pi T}g_{(0)}^{ij}v_j \partial_{i}a_{t}^{(0)}\right) \nn
&+\sqrt{g_{(0)}}Z^{(0)}e^{-i\omega v_{EF}}a^{(0)}_{t}\,\left( e^{i\omega v_{EF}}+\frac{1}{2}\left(\delta g^{(0)}_{tt}-\delta g^{(0)}_{rr}+g_{(0)}^{ij} \delta g^{(0)}_{ij}\right) \right)\nn
&+\sqrt{g_{(0)}}\,\partial_{\phi}Z^{(0)}e^{-i\omega v_{EF}}a^{(0)}_{t}\,\delta\phi^{(0)}\,.
\end{align}

We can now evaluate the constraints at the horizon. Expanding the Gauss law constraint $C=\partial_{\mu}\pi^{\mu}=0$ gives
\begin{align}\label{cconstraint}
&\nabla_i\left( Z^{(0)} \left( \nabla^{i}w +i\omega g_{(0)}^{ij}\delta a^{(0)}_{j} +v^{i}a_{t}^{(0)}\right) \right) = i\omega Z^{(0)}\frac{1}{2}a^{(0)}_{t}\,\left(\delta g^{(0)}_{tt}-\delta g^{(0)}_{rr}+g_{(0)}^{ij} \delta g^{(0)}_{ij}\right) \nn
&+i\omega Z^{(0)} \left( \delta a^{(1)}_{t}-\frac{i \omega}{4\pi T}\left( \delta a^{(1)}_{t}-\delta a^{(1)}_{r}  \right) + \frac{1}{4\pi T}v^i \partial_{i}a_{t}^{(0)}\right)+i\omega\,\partial_{\phi}Z^{(0)}a^{(0)}_{t}\,\delta\phi^{(0)}\,.
\end{align}
To expand the momentum constraints at the horizon, $H_{\nu}=0$, we first note that
\begin{align}\label{fpimuathor}
f_{t\mu}\pi^{\mu}&=F_{ti}\pi^{i} \rightarrow 0\,,\nn
f_{i\mu}\pi^{\mu}&=F_{it}\pi^{t}+F_{ij}\pi^{j} \rightarrow \sqrt{g_{(0)}}Z^{(0)}e^{-i\omega v_{EF}}a^{(0)}_{t}\,\left( \partial_{i}w+i\omega\delta a_{i}^{(0)} \right)\,,
\end{align}
For the $t$ component, $H_{t}=0$, we then find
\begin{align}\label{htconstraint}
\left(2\,\nabla_{i}v^{i}-i\omega g_{(0)}^{ij}\delta g^{(0)}_{ij} \right)\left(1+\frac{i\omega}{2\pi T}\right) =0 \,.
\end{align}
Similarly for the 
$i$ component $H_{i}=0$ we get 
\begin{align}\label{hiconstraint}
&-2\,\nabla^{j}\nabla_{(j}v_{i)} +\nabla_{i}p^{\prime}+i\omega\left(-\delta g_{ti}^{(1)}+\partial_{i}\delta g_{tr}^{(0)}-g^{(1)}_{il}v^{l}+\frac{i\,\omega}{4\pi T}(\delta g_{ti}^{(1)}-\delta g_{ri}^{(1)})+\nabla^{k}\delta g^{(0)}_{ki} \right) \nn
&-Z^{(0)}a^{(0)}_{t}\,\left( \nabla_{i}w+i\omega\delta a_{i}^{(0)} \right) +\nabla_{i}\phi^{(0)} 
\left( v^{j} \nabla_{j}\phi^{(0)}-i\omega \delta\phi^{(0)}\right)
=0\,.
\end{align}
Finally, we consider the Hamiltonian constraint $\left(-h\right)^{-1/2}H=0$. The third and fifth terms in \eqref{constraints}
vanish at linearised order. The second and the sixth term is of order $\mathcal{O}(r^{0})$. We also compute 
\begin{align}\label{flclo}
f_{ri}&\rightarrow -\frac{i\omega e^{-i\omega v_{EF}}}{4\pi T r}\,\delta a^{(0)}_{i}\,+{\cal{O}}(r^{0})\,,\nn
f_{rt}&\rightarrow -\frac{i\omega e^{-i\omega v_{EF}}}{4\pi T r}\,w\,+{\cal{O}}(r^{0})\,,\nn
f_{ti}&\rightarrow F_{ti}\,,\nn
f_{ij}&\rightarrow F_{ij}\,,
\end{align}
and so the fourth term is of order $\mathcal{O}(r^{0})$ as well. Finally, the first term turns out to be of order $\mathcal{O}(r^{-1})$, leading to the constraint
\begin{align}\label{hconstraint}
2\,\nabla_{i}v^{i}-i\omega g_{(0)}^{ij}\delta g^{(0)}_{ij}=0 \,,
\end{align}
which is consistent with \eqref{htconstraint}.
Finally, after using  \eqref{eq:nh_exp3} and \eqref{hconstraint} in \eqref{hiconstraint}, we find that the latter takes the form
\begin{align}\label{hiconstraint2}
&i\omega\left(-\delta g_{ti}^{(1)}-g^{(1)}_{il}v^{l}+\partial_{i}(\delta g_{tr}^{(0)}-\delta g_{rr}^{(0)}) +\frac{i\,\omega}{4\pi T}(\delta g_{ti}^{(1)}-\delta g_{ri}^{(1)})+\nabla^{k}\delta g^{(0)}_{ki} \right) \nn
&-2\,\nabla^{j}\nabla_{(j}v_{i)}+\nabla_{i}p -Z^{(0)}a^{(0)}_{t}\,\left( \nabla_{i}w+i\omega\delta a_{i}^{(0)} \right) +\nabla_{i}\phi^{(0)} 
\left( v^{j} \nabla_{j}\phi^{(0)}-i\omega \delta\phi^{(0)}\right) =0\,,
\end{align}
with
\begin{align}\label{eq:p_def}
p=-2\pi T(\delta g_{tt}^{(0)}+\delta g_{rr}^{(0)})\,.
\end{align}

Thus, in summary, equations \eqref{cconstraint},\eqref{hconstraint} and \eqref{hiconstraint2} are the constraint equations for the perturbations on the horizon.

\section{Calculating the DC conductivity}\label{dcapp}
We briefly summarise the results of \cite{Donos:2015gia,Banks:2015wha} which allows us to obtain a horizon 
DC conductivity by solving a system of Stokes equations on the horizon. When the DC conductivity of the dual field theory is finite,
as in the case of explicit breaking of translations, it is identical to the horizon DC conductivity.

By analysing a perturbation of the background black hole solutions \eqref{eq:DC_ansatz} that, crucially, incorporate
DC sources, it was shown that one is led to the following system of Stokes equations on the black hole horizon
\begin{align}\label{DCone}
\partial_{i} Q^{i}_{(0)}=0\,, \qquad
\partial_{i} J^{i}_{(0)}=0\,,\nn
-2\,\nabla^{i}\nabla_{\left( i \right. }v_{\left. j\right)}-{Z^{(0)}a_{t}^{(0)}}\nabla_j w
+\nabla_j\phi^{(0)}\nabla_i\phi^{(0)}v^{i}
+\nabla_{j}\,p&=4\pi T\,\bar \zeta_{j}
+{Z^{(0)}a_{t}^{(0)}} \bar E_j\,,
\end{align}
where
\begin{align}\label{eq:J_hor2ap}
Q^i_{(0)} &=4\pi T\sqrt{g_{(0)}}v^j\,,\nn
J^i_{(0)}&=\sqrt{g_{(0)}}g^{ij}_{(0)}Z^{(0)}\left(\partial_j w+{a^{(0)}_t}v_j+\bar E_j\right)\,.
\end{align}
Here, the vectors $\bar E_i$,$\bar\zeta_i$, which are taken to be constant, 
parametrise the DC electric source and thermal gradient of the dual field theory, respectively.

After solving these Stokes equations we obtain the local currents $Q^i_{(0)}$, $J^i_{(0)}$ on the horizon as functions of the DC sources $\bar E_i,\bar\zeta_i$. By defining the current flux densities at the horizon via
\begin{align}
\bar J^i_{(0)}=\int J^i_{(0)}\,,\qquad \bar Q^i_{(0)}=\int Q^i_{(0)}\,,\qquad 
\end{align}
we can then define the horizon thermoelectric DC conductivity matrix via
\begin{align}\label{bigform2}
\left(
\begin{array}{c}
\bar J^i_{(0)}\\\bar Q_{(0)}^i
\end{array}
\right)=
\left(\begin{array}{cc}
\sigma_H^{ij} & T\alpha_H^{ij} \\
T\bar\alpha_H^{ij} &T\bar\kappa_H^{ij}   \\
\end{array}\right)
\left(
\begin{array}{c}
\bar E_j\\ \bar\zeta_j
\end{array}
\right)\,.
\end{align}
For the time reversal invariant backgrounds we are considering in this paper we have $\sigma_H$, $\bar\kappa_H$ are symmetric and $\alpha_H=\bar\alpha_H^T$. 

Furthermore, as explained in \cite{Donos:2015gia,Banks:2015wha}, the current flux densities at the horizon, defined by
\begin{align}
\bar J^i_{(0)}=\int J^i_{(0)}\,,\qquad \bar Q^i_{(0)}=\int Q^i_{(0)}\,,\qquad 
\end{align}
are identical to the current fluxes $\bar J^i$, $\bar Q^i$ of the dual field theory. Thus, for holographic lattices we have
the DC conductivities of the dual field theory, 
$\sigma$, $\alpha$, $\bar\alpha$ and $\bar\kappa$,
are identical to the horizon conductivities
$\sigma_H$, $\alpha_H$, $\bar\alpha_H$ and $\bar\kappa_H$, respectively.

It is helpful for the analysis of this paper to recall from \cite{Donos:2015gia,Banks:2015wha} that as long as the horizon does not have any Killing vectors, there is a unique solution to the Stokes equations \eqref{DCone}, up to undetermined constants in $w$ and $p$, which do not inhibit one solving for
the DC conductivity since they do not enter the expressions for the currents.

Finally, we emphasise that the horizon DC conductivity given in \eqref{bigform2} should not be confused with another notion of horizon conductivity that arises from the constitutive relations for the auxiliary fluid on the horizon. For example, in the expression for the 
electric current on the horizon given in \eqref{eq:J_hor2ap}, one can call $\sqrt{g_{(0)}}g^{ij}_{(0)}Z^{(0)}$ a local
electric conductivity\footnote{To avoid confusion, we note that in \cite{Donos:2015gia} the expression $\sqrt{g_{(0)}}g^{ij}_{(0)}Z^{(0)}$ was denoted by $\sigma^{ij}_H$, a notation which we do not use here.}
, but this is, in general quite distinct from $\sigma_H^{ij}$ as defined in \eqref{bigform2}.

\section{Counting functions of integration}\label{contex}

The quasinormal diffusion modes are solutions to the bulk equations of motion satisfying ingoing boundary conditions
at the black hole horizon and have vanishing source terms at the AdS boundary. In the text we focused on
the constraint equations given in section \ref{coneqsb} as this was sufficient to extract the diffusion relation for the modes.
Here, and in the next appendix, we outline how, with the time dependence given in \eqref{peeeq}, we have specified enough 
data\footnote{We note that the arguments will need to be modified for models in which the holographic lattices
have additional Goldstone modes due, for example, to
the breaking of an internal abelian symmetry. The presence of moduli in the bulk solution, arising from a global symmetry in the bulk, for example,
will also lead to additional modifications.} at the 
horizon and at the AdS boundary in order to obtain a solution to the full equations of motion. 

We begin with fluctuations of the scalar field which satisfies a second order equation in the radial variable. At the $AdS$ boundary, $r\to \infty$, 
we have two functions of integration, depending on the spatial coordinates $x^i$, associated with the source terms and expectation value of
the scalar operator in the dual CFT. At the black hole horizon we demand that the perturbation is regular and this leaves us with a single function
of integration $\delta\phi^{(0)}(x)$. By setting the source terms at the AdS boundary to zero we can
develop a solution in the bulk using the remaining function at the AdS boundary and then match with the solution developed from the horizon using
$\delta\phi^{(0)}(x)$ which leads, generically, to a unique solution everywhere.

We next turn to fluctuations of the gauge field. The radial component $\delta a_r$ serves as a Lagrange multiplier and is data which, {\it a priori}, we are free to specify. 
This leaves
$D-1$ functions $\delta a_i$, $\delta a_t$, each of which satisfies a second order differential equation in the radial variable, and there
is also the Gauss constraint, $C=0$ (see \eqref{constraints}), that we impose infinitesimally close to the horizon, given in \eqref{hconstrainttwo}. 
At the AdS boundary we set the $D-1$ functions of integration
that are associated with possible source terms to zero, implying that we need to identify $(D-1)$ functions from the horizon expansion in order to
solve the second order equations of motion, via a matching argument. 
With the ingoing boundary conditions \eqref{eq:nh_exp2},\eqref{eq:nh_exp3}
at the horizon, we have 
the functions $\delta a_{t}^{(0)}$ and, when $\omega\ne 0$, $\delta a_{i}^{(0)}$, $\delta a_{t}^{(1)}$ all appearing in the constraint equation. 
If we pick $\delta a_{t}^{(0)}$ to be solved by the constraint equation then we are left
with precisely $D-1$ functions $\delta a_{i}^{(0)}$ and $\delta a_{t}^{(1)}$ which are fixed by the matching. 
It is worth noting that in our procedure, for the leading term of the Lagrange multiplier we must set $\delta a^{(0)}_r=w$ (see \eqref{eq:nh_exp3}), 
which we are free to do. In addition, we note that $\delta a^{(1)}_r$, the sub-leading term of the Lagrange multiplier, also appears in \eqref{hconstrainttwo} and can be chosen freely; in particular
$\delta a^{(1)}_r$ does not affect the in-falling conditions we have specified in \eqref{eq:nh_exp2}, \eqref{eq:nh_exp3}.

Finally, we discuss the metric fluctuations, which 
run along similar lines to the gauge field. There are $D(D+1)/2$ metric functions $\delta g_{\mu\nu}$ out of which the $D$ functions
$\delta g_{r\mu}$ serve as Lagrange multipliers. The remaining $D(D-1)/2$ functions, $\delta g_{tt}$, $\delta g_{ti}$ and
$\delta g_{ij}$, each satisfy differential equations which are second order in the radial direction. There are also $D$ constraint equations, the Hamiltonian and momentum constraints, $H=H^\nu=0$ (see \eqref{constraints}),
that we have chosen to impose on a surface infinitesimally close to the horizon. Importantly,
however, the Hamiltonian constraint is redundant close to the horizon leaving $D-1$ independent constraint equations to satisfy near the horizon,
given in \eqref{hconstrainta} and \eqref{hiconstraint2b}. There are now two equivalent ways to proceed, 
which we discuss in turn.

The first way is to solve the second order radial equations for $\delta g_{tt}$, $\delta g_{ti}$ and
$\delta g_{ij}$. 
Now with the ingoing boundary conditions \eqref{eq:nh_exp1}, \eqref{eq:nh_exp3} these equations are associated with $D(D-1)/2+(D-2)$ functions of integration on the horizon, $\delta g_{tt}^{(0)}$, $\delta g_{ij}^{(0)}$, $\delta g_{ti}^{(1)}$, and $v_{i}$, all of which appear in the constraint equations when $\omega\ne 0$ and we are ignoring the functions associated with the Lagrange multipliers appearing at the horizon for the moment.
Close to the AdS boundary, where we fix the source terms to zero, the second order equations give a set of $D(D-1)/2$ of normalisable modes that will be fixed along with the $D(D-1)/2$ functions $\delta g_{tt}^{(0)}$, $\delta g_{ij}^{(0)}$, $\delta g_{ti}^{(1)}$ upon matching in the bulk. The remaining $D-1$ functions on the horizon, $v_{i}$ and $p$, are then used to solve the momentum constraints $H^{\nu}=0$. The issue that remains open in this approach is the Hamiltonian constraint. One can show that in an expansion close to the horizon this is satisfied order by order in an analytic radial expansion provided that
the remaining constraint equations and second order equations are satisfied, along with imposing
the boundary conditions \eqref{eq:nh_exp1}, \eqref{eq:nh_exp2} and \eqref{eq:nh_exp3}. Away from the horizon, this is guaranteed by the fact that $\mathcal{L}_r{H}=0$.

The second way is to solve the Hamiltonian constraint equation in the bulk instead of the second order equation for $\delta g_{tt}$. Indeed, one of the second order radial equations, for example the one for $\delta g_{tt}$, is implied by the Hamiltonian constraint. This is because the Hamiltonian constraint contains no derivatives of the momentum and hence only $\partial_{r} \delta g_{tt}$ appears, along with second order spatial derivatives with respect to $x^i$. Thus, instead of the $D(D-1)/2$ second order equations in the radial direction, we just need to solve $D(D-1)/2-1$ second order equations and one first order equation. Setting the source terms to zero in these equations at the AdS boundary, we conclude that we need to specify $D(D-1)/2-1+1$ free functions at the horizon after imposing the ingoing boundary conditions and solving the $D-1$ constraint equations \eqref{hconstrainta},\eqref{hiconstraint2b}. If we again use the $D-1$ constraint equations to solve for $v_i$, $p$, this will leave precisely $D(D-1)/2$ functions $\delta g_{ij}^{(0)}$, $\delta g_{tt}^{(0)}$, $\delta g_{ti}^{(1)}$ to be fixed by the matching.

Concerning the Lagrange multipliers, we first note that in the above procedure $\delta g_{rt}^{(0)}\equiv -p/(4\pi T)$ will be fixed.
Furthermore, we must set $\delta g_{ri}^{(0)}=-v_{i}$, $\delta g_{rr}^{(0)}=-p/(2\pi T)-\delta g_{tt}^{(0)}$ (see \eqref{eq:nh_exp3}).
In addition, we also note that the sub-leading term $\delta g^{(1)}_{ri}$ appears in \eqref{hiconstraint2b} and can be chosen freely as part of fixing the Lagrange multipliers. Notice that, similarly to the case of the radial component of the gauge field $\delta a_{r}^{(1)}$, this does not spoil the in-falling conditions we have specified in \eqref{eq:nh_exp1}, \eqref{eq:nh_exp3}.

\section{Fixing the zero modes of the $\varepsilon$ expansion}\label{app:expansion}
We now examine the perturbation at third order in $\varepsilon$. It is useful to split all the bulk fields $\mathbf{\Phi}_{[i]}$ according to
\begin{align}\label{eq:app_pertb_split}
\mathbf{\Phi}_{[i]}=\mathbf{\hat{\Phi}}_{[i]}+\frac{\partial \mathbf{\Phi}_{b}}{\partial T}\,\delta T_{[i]}+\frac{\partial \mathbf{\Phi}_{b}}{\partial \bar{\mu}}\,\delta \bar{\mu}_{[i]}\,.
\end{align}
Here the second and third terms are the derivatives of the background solution with respect to 
the temperature and the averaged chemical potential, in the gauge described below \eqref{nhpeth}; in particular at the horizon
the derivatives are explicitly given in eq. \eqref{seedqnm}. The functions $\hat{\mathbf{\Phi}}_{[i]}$ solve the perturbative in $\varepsilon$ radial equations of motion with boundary conditions on the horizon set by $\hat{w}_{[i]}$, $\hat{p}_{[i]}$ and $v^{j}_{[i]}$ which are obtained from the perturbatively expanded horizon constraint equations and
we recall that 
$\hat{w}_{[i]}$, $\hat{p}_{[i]}$ do not have a zero mode (see \eqref{eq:hmode_split}). We stress that this doesn't necessarily mean that the bulk functions $\hat{\mathbf{\Phi}}$ do not have a zero mode. However, when $\mathbf{\Phi}$ is the field $\delta a_{t}$, for example, our boundary conditions impose that $\Hat{(\delta a_{t})}_{[i]}$ is equal to $\hat{w}_{[i]}$ on the horizon and that function does not have a zero mode.

In this notation, at third order, the scalar constraint equations \eqref{hconstrainta} and \eqref{hconstrainttwo} read 
\begin{align}\label{thorderce}
&\nabla_{i}v_{[3]}^{i}=\frac{i\omega_{[3]}}{2}\left(g_{(0)}^{ij} \frac{\partial g^{(0)}_{ij}}{\partial T} \,\delta T+ g_{(0)}^{ij} \frac{\partial g^{(0)}_{ij}}{\partial \bar \mu}\,\delta \bar \mu\right)\nn
&\qquad\qquad +\frac{i\omega_{[2]}}{2}\left(g_{(0)}^{ij} \frac{\partial g^{(0)}_{ij}}{\partial T}\,\delta T_{[1]} +g_{(0)}^{ij} \frac{\partial g^{(0)}_{ij}}{\partial \bar \mu}\,\delta \bar \mu_{[1]} \right) +\frac{i\omega_{[2]}}{2}g^{ij}_{(0)}\delta \hat{g}^{(0)}_{[1]ij}-ik_{i}v_{[2]}^{i}\,,
\end{align}
and
\begin{align}\label{thorderce2}
&\nabla_j\left( Z^{(0)} \left(i\,k^{j}\delta\bar{\mu}_{[2]}+ \nabla^{j}w_{[3]} +v^{j}_{[3]}a_{t}^{(0)}\right) \right)=\nn
&-\nabla_j\left( Z^{(0)} \left(i\,k^{j}\hat{w}_{[2]} +i\omega_{[2]}g^{jk}_{(0)}\delta \hat{a}^{(0)}_{[1]k}\right) \right) - ik_{j}\left( Z^{(0)} \left(i\,k^{j}(\delta\bar{\mu}_{[1]}+\hat{w}_{[1]})+ \nabla^{j}\hat{w}_{[2]} +v^{j}_{[2]}a_{t}^{(0)}\right) \right)\nn
& +i\omega_{[3]} \,\left(\frac{1}{2}Z^{(0)}a^{(0)}_{t}g_{(0)}^{ij}  \frac{\partial g^{(0)}_{ij}}{\partial T}+\partial_{\phi} Z^{(0)}a^{(0)}_{t}\, \frac{\partial \phi^{(0)}}{\partial T}+Z^{(0)} \frac{\partial a_{t}^{(0)}}{\partial T}\right)\delta T\nn
& +i\omega_{[3]} \,\left(\frac{1}{2}Z^{(0)}a^{(0)}_{t}g_{(0)}^{ij} \frac{\partial g^{(0)}_{ij}}{\partial \bar \mu}+\partial_{\phi} Z^{(0)}a^{(0)}_{t}\,\frac{\partial \phi^{(0)}}{\partial \bar \mu}+Z^{(0)}\frac{\partial a_{t}^{(0)}}{\partial\bar \mu}\right)\delta\bar\mu\nn
&+ i\omega_{[2]} \,\left(\frac{1}{2}Z^{(0)}a^{(0)}_{t}g_{(0)}^{ij}  \frac{\partial g^{(0)}_{ij}}{\partial T}+\partial_{\phi} Z^{(0)}a^{(0)}_{t}\, \frac{\partial \phi^{(0)}}{\partial T}+Z^{(0)} \frac{\partial a_{t}^{(0)}}{\partial T}\right)\delta T_{[1]}\nn
& +i\omega_{[2]} \,\left(\frac{1}{2}Z^{(0)}a^{(0)}_{t}g_{(0)}^{ij} \frac{\partial g^{(0)}_{ij}}{\partial \bar \mu}+\partial_{\phi} Z^{(0)}a^{(0)}_{t}\,\frac{\partial \phi^{(0)}}{\partial \bar \mu}+Z^{(0)}\frac{\partial a_{t}^{(0)}}{\partial\bar \mu}\right)\delta\bar\mu_{[1]}\nn
&+i\omega_{[2]} \,Z^{(0)}\frac{1}{2}a^{(0)}_{t}\,\left(\delta \hat{g}^{(0)}_{[1]tt}-\delta \hat{g}^{(0)}_{[1]rr}+g_{(0)}^{ij} \delta \hat{g}^{(0)}_{[1]ij}\right) \nn
&+i\omega_{[2]} \,Z^{(0)} \left( \delta \hat{a}^{(1)}_{[1]t} + \frac{1}{4\pi T}v^{i}_{[1]} \partial_{i}a_{t}^{(0)}\right)+i\omega_{[2]}\,\partial_{\phi}Z^{(0)}a^{(0)}_{t}\,\delta\hat{\phi}^{(0)}_{[1]}\,,
\end{align}
respectively, and we recall $\omega_{[1]}=0$. We will not explicitly write out the vector constraint equation \eqref{hiconstraint2b} at this order, which is rather long, and in fact won't play a role in the following discussion.
We now want to examine the global constraints implied by the requirement that the periodic functions $v_{[3]}^{i}$, $\hat{w}_{[3]}$ and $\hat{p}_{[3]}$ exist. After integrating the two equations \eqref{thorderce},\eqref{thorderce2} over space we obtain an inhomogeneous system of algebraic equations involving the constants $\delta T_{[1]}$, $\delta\hat{\mu}_{[1]}$, which are, so far, undetermined,
as well as $\omega_{[3]}$. As we will now discuss, the parts of this system that are not homogeneous in these variables
will involve integrals of functions which are fixed at first order in perturbation theory, as well as $\omega_{[2]}$ 
which was already fixed in the main text, following \eqref{modeseqn}. 
As we will see, it is important to identify the
implicit dependence of these two equations on $\delta T_{[1]}$ and $\delta\bar{\mu}_{[1]}$ as well as the manifest explicit dependence.

After solving the leading order constraint equations as well as the radial equations, we know that
$\hat{w}_{[1]}$, $\hat{p}_{[1]}$,$v_{[1]}^{i}$ and indeed
all the first order functions $\hat{\mathbf{\Phi}}_{[1]}$
are proportional to the constants $\delta T$ and $\delta\mu$. Furthermore, the constants $\delta T$ and $\delta\mu$ are not independent of each other: from \eqref{modeseqn}-\eqref{twoevals}, the existence of the perturbation at second order imposes that for each of the two diffusive modes we must have
\begin{align}
\left(\begin{array}{c}
\delta T_{\pm}\\\delta\bar{\mu}_{\pm}\,
\end{array}
\right)=\delta h\mathbb{V}_{\pm}\,,
\end{align}
where $\delta h$ is a constant, the vector $\mathbb{V}_{\pm}$  belongs to the kernel of the matrix
\begin{align}
\mathbb{M}_{\pm}=\left(\begin{array}{cc}
T^{-1}\left(i\,\omega_{[2]}^{\pm}\,c_{\mu}-\bar{\kappa}(k)\right) & i\,\omega_{[2]}^{\pm}\,\xi-\alpha(k)\\
i\,\omega_{[2]}^{\pm}\,\xi-\alpha(k) & i\,\omega_{[2]}^{\pm}\,\chi-\sigma(k)
\end{array}
\right)\,,
\end{align}
and $\omega^\pm_{[2]}$ is given in \eqref{twoevals}.
It is also convenient to introduce the vector $\mathbb{V}^{\perp}_{\pm}$ which is orthogonal to $\mathbb{V}_{\pm}$. Then there is some constants $\delta h^{||}_{[1]}$ and $\delta h^{\perp}_{[1]}$ such that we can write
\begin{align}
\left(\begin{array}{c}
\delta T_{[1]}\\\delta\bar{\mu}_{[1]}
\end{array}
\right)=\delta h^{||}_{[1]}\,\mathbb{V}_{\pm}+\delta h^{\perp}_{[1]}\,\mathbb{V}^{\perp}_{\pm}\,.
\end{align}
We stress here that the component $\delta h^{||}_{[1]}$ is redundant and we do not expect it to be fixed by the equations of motion. This follows from the fact we are examining a linearised perturbation and we should be able to freely choose $\delta h^{||}_{[1]}$ by scaling the whole solution by a function of $\varepsilon$. Multiplying the whole perturbation by e.g. $1+\alpha\,\varepsilon$ would effectively shift $\delta h^{||}_{[1]}\to \delta h^{||}_{[1]}+\alpha\,\delta h$. As we will see, the constant $\delta h^{||}_{[1]}$ does indeed drop out from the two algebraic equations that we obtain by integrating over \eqref{thorderce},\eqref{thorderce2}.

Using this notation, for the first order functions $\hat{\mathbf{\Phi}}_{[1]}$, which just depend on $\delta T$,$\delta\mu$ we can write
\begin{align}
\hat{\mathbf{\Phi}}_{[1]}&=\hat{\mathbf{\Phi}}_{[1][0]}\,\delta h\,.
\end{align}
For the second order functions $\hat{\mathbf{\Phi}}_{[2]}$ the situation is more involved and we have
\begin{align}
\hat{\mathbf{\Phi}}_{[2]}&=\hat{\mathbf{\Phi}}_{[2][0]}\,\delta h+\hat{\mathbf{\Phi}}_{[2]T}\,\delta T_{[1]}+\hat{\mathbf{\Phi}}_{[2]\mu}\,\delta \bar{\mu}_{[1]}\,,\nn
&=\hat{\mathbf{\Phi}}_{[2][0]}\,\delta h+\hat{\mathbf{\Phi}}_{[2][1]}\,\delta h^{||}_{[1]}+\hat{\mathbf{\Phi}}^{\perp}_{[2][1]}\,\delta h^{\perp}_{[1]}\,.
\end{align}
In particular
\begin{align}\label{secopertap}
\hat{w}_{[2]}&=\hat{w}_{[2][0]}\delta h+\hat{w}_{[2]T}\,\delta T_{[1]}+\hat{w}_{[2]\mu}\,\delta \bar{\mu}_{[1]}\,,\nn
\hat{p}_{[2]}&=\hat{p}_{[2][0]}\delta h+\hat{p}_{[2]T}\,\delta T_{[1]}+\hat{p}_{[2]\mu}\,\delta \bar{\mu}_{[1]}\,,\nn
v^{i}_{[2]}&=v^{i}_{[2][0]}\delta h+v^{i}_{[2]T}\,\delta T_{[1]}+v^{i}_{[2]\mu}\,\delta \bar{\mu}_{[1]}\,.
\end{align}
The parts of the solutions of these horizon quantities that are proportional to $\delta h$ can be found from the constraints \eqref{secorderce} after setting $\delta T_{[1]}$ and $\delta\bar{\mu}_{[1]}$ equal to zero. 
The key observation, now,  is that in the constraint equations at second order (i.e. \eqref{secorderce} as well as the vector constraint equation),
the pieces in \eqref{secopertap} proportional to $\delta T_{[1]}$ and $\delta \bar{\mu}_{[1]}$ are precisely the same equations that we have in 
the DC calculation outlined in Appendix \ref{dcapp} with $E_{i}=-i\,k_{i}\,\delta\bar{\mu}_{[1]}$ and $\zeta_{i}=-i\,k_{i}\,\delta T_{[1]}/T$. We can therefore write
\begin{align}
&4\pi T i \oint \sqrt{g_{(0)}}v_{[2]T}^{i}=\bar\kappa^{ij}_Hk_j\,,\qquad
4\pi T i \oint \sqrt{g_{(0)}}v_{[2]\mu}^{i}=T\bar\alpha^{ij}_Hk_j\,,\nn
&i\oint \sqrt{g_{(0)}}Z^{(0)} \left(-i\,k^{i} + \nabla^{i}\hat{w}_{[2]\mu} +v_{[2]\mu}^{i}a_{t}^{(0)}\right) =\sigma^{ij}_Hk_j\,,\nn
&i\oint \sqrt{g_{(0)}}Z^{(0)} \left(\nabla^{i}\hat{w}_{[2]T} +v_{[2]T}^{i}a_{t}^{(0)}\right) =\alpha^{ij}_Hk_j\,.
\end{align}

With the ingredients assembled above, we now integrate equations \eqref{thorderce},\eqref{thorderce2} and find that we can write them in the form
\begin{align}\label{eq:third_order_fixing}
i\,\omega_{[3]}\,\mathbb{S}\mathbb{V}_{\pm}\delta h+\mathbb{M}_{\pm}\left(\mathbb{V}_{\pm}\,\delta h^{||}_{[1]}+\mathbb{V}^{\perp}_{\pm}\,\delta h^{\perp}_{[1]}\right)+\mathbb{W}\,\delta h&=0\,,\nn
\Rightarrow i\,\omega_{[3]}\,\mathbb{S}\mathbb{V}_{\pm}\delta h+\mathbb{M}_{\pm}\,\mathbb{V}^{\perp}_{\pm}\,\delta h^{\perp}_{[1]}+\mathbb{W}\,\delta h&=0\,,
\end{align}
where we have defined the matrix of susceptibilities
\begin{align}
\mathbb{S}=\left( \begin{array}{cc}
T^{-1}c_{\mu} &\xi\\
\xi &\chi
\end{array} \right)\,.
\end{align}
The vector $\mathbb{W}\delta h$ is defined through the integrals of the {\it functions} that appear in \eqref{thorderce},\eqref{thorderce2} with index $[1]$ and also through the $v^{i}_{[2][0]}$ part of the horizon fluid velocity, both of which are proportional to $\delta h$. Equation \eqref{eq:third_order_fixing} now fixes both $\omega_{[3]}$ as a function of $k^{i}$ and $\delta h_{[1]}^{\perp}$ as a function of $\delta h$. In particular, this shows how the zero modes of $\delta T_{[1]}$, $\delta\bar{\mu}_{[1]}$ are fixed at this order of perturbation theory. 

It is also clear from the above analysis, that a similar structure will repeat itself at higher orders in the perturbative expansion, fixing
the zero modes of $\delta T_{[i]}$, $\delta\bar{\mu}_{[i]}$ for $i>1$. 
In particular, in the expression \eqref{eq:app_pertb_split} we will have $\mathbf{\hat{\Phi}}_{[i]}$ depending on
$\delta T_{[i-1]}$ and $\delta \bar{\mu}_{[i-1]}$
in an analogous way.


\providecommand{\href}[2]{#2}\begingroup\raggedright\endgroup

\end{document}